\documentclass[journal=jacsat,manuscript=article]{achemso}
\usepackage{ulem}

\usepackage[version=3]{mhchem} % Formula subscripts using \ce{}

\usepackage[justification=raggedright]{caption}
\usepackage{physics}
\usepackage{graphicx}
\usepackage{subcaption}
\usepackage{dcolumn}
\usepackage{amsmath}
\usepackage{amssymb}
\usepackage{color}
\usepackage{multirow}
\usepackage{booktabs}
\usepackage{tabularx}
\usepackage{rotating}
\usepackage{adjustbox}
\usepackage{hyperref}
\usepackage{cleveref}
\usepackage{nameref}
\usepackage{footnote}
\usepackage{verbatim,moreverb,bm}
\def\up{\uparrow}
\def\down{\downarrow}
\def\bef{\begin{framed}}
\def\eef{\end{framed}}
\def\be{\begin{equation}}
\def\ee{\end{equation}}
\def\ber{\begin{eqnarray}}
\def\eer{\end{eqnarray}}
\def\xiv{{\boldmath{\xi}}}
\def\zetav{{\boldmath{\zeta}}}

\def\nablabold{\mbox{\boldmath $\nabla$}}

\def\rv{{\bf r}}

\def\Qv{{\bf Q}}

\def\xv{{\bf \hat x}}
\def\yv{{\bf \hat y}}
\def\zv{{\bf \hat z}}

\def\kv{{\bf k}}
\def\qv{{\bf q}}

\def\Cv{{\bf C}}

\def\Gv{{\bf G}}

\def\Rv{{\bf R}}
\def\Qv{{\bf Q}}
\def\Xv{{\bf X}}

\def\nn{\nonumber}
\def\xiv{\bm{\xi}}
\def\zetav{\bm{\zeta}}
\def\gammav{\bm{\gamma}}

\SectionNumbersOn

\title{Orbital-Free Density Functional Theory for Periodic Solids: \\ Construction of the Pauli Potential}

\author{Sangita Majumdar}
\affiliation{Institute for Functional Intelligent Materials (I-FIM), National University of Singapore, 4 Science Drive 2, Singapore 117544}

\author{Zekun Shi}
\affiliation{SEA AI Lab, Singapore}
\alsoaffiliation{School of Computing, National University of Singapore, 13 Computing Drive, Singapore 117417}

\author{Giovanni Vignale}
\affiliation{Institute for Functional Intelligent Materials (I-FIM), National University of Singapore, 4 Science Drive 2, Singapore 117544}
\email{vgnl.g@nus.edu.sg}

\begin{document}

\begin{abstract}
The practical success of density functional theory (DFT) is largely credited to the Kohn-Sham approach, which enables the exact calculation of
the non-interacting electron kinetic energy via an auxiliary noninteracting system. Yet, the realization of DFT's full potential awaits
the discovery of a direct link between the electron density, $n$, and the non-interacting kinetic energy, $T_{S}[n]$.
In this work, we address two key challenges towards this objective. First, we introduce a new algorithm for directly solving the
constrained minimization problem yielding $T_{S}[n]$ for periodic
densities -- a class of densities that, in spite of its central importance for materials science, has received limited attention in the literature. Second, we present a numerical procedure that allows us to calculate the functional derivative of $T_{S}[n]$ with respect to the density at constant electron number,  also known as the Kohn-Sham potential $V_{S}[n](\rv)$. Lastly, the algorithm is augmented with a subroutine that computes the ``derivative discontinuity", i.e., the spatially uniform jump in $V_{S}[n](\rv)$  which occurs upon increasing or decreasing the total number of electrons. This feature allows us to distinguish between ``insulating" and ``conducting" densities for non interacting electrons.
The code integrates key methodological innovations, such as the use of an adaptive basis set (``equidensity orbitals") for wave function expansion and the QR decomposition to accelerate the implementation of the orthogonality constraint. Notably, we derive a closed-form expression for the Pauli potential in one dimension, expressed solely in terms of the input density, without relying on Kohn-Sham eigenvalues and eigenfunctions. We validate this method on one-dimensional periodic densities, achieving results within ``chemical accuracy".
\end{abstract}
\maketitle

\section{Introduction}
The original density functional theory (DFT) of Hohenberg and Kohn \cite{hohenberg64}  purported to replace the variational principle of quantum mechanics, in which the energy is minimized with respect to a many-body wave function, by a much simpler variational principle in which the energy is minimized with respect to the electronic density, a function of position denoted by $n(\rv)$.
In practice, that bold vision never materialized. The second landmark paper in the field (Kohn-Sham, KS) \cite{kohn1965} already introduced a representation of the electronic density in terms of orthogonal single-particle orbitals (still much simpler than the full many-body wave function) and the minimization of the energy with respect to the density was replaced by the solution of self-consistent equations for the single-particle orbitals, which are now known as the Kohn-Sham equations.  By doing that, they not only ensured that the electronic density would be compatible with the requirements of the Pauli exclusion principle, but also that one of the largest components of the energy, the non-interacting kinetic energy, defined precisely below, would be treated exactly. From then onwards, much of the research in DFT focused on devising good approximations for the genuinely many-body part of the energy functional, the exchange-correlation energy functional $E_{xc}[n]$ and its functional derivative with respect to density, known as the exchange-correlation potential,  $V_{xc}[n](\rv)$. While tremendous progress was made in this direction, many problems that lurked under the smooth surface of the formalism gradually came into focus (self-interaction error, derivative discontinuity, band gaps, strong correlations, degenerate ground states,  broken symmetry) \cite{cohen2012, verma2020} and the community learned to cope with them with varying degrees of success.

However, the quest for a practical orbital-free implementation of DFT was never abandoned \cite{karasiev2012issues}. At the heart of the quest lies the formidable problem of constructing the universal energy functional $F[n]$ (sometimes referred to as the ``Holy Grail" of DFT) \cite{lieb1983, perdew1981,lewin2023} whose formal definition is [Levy-Lieb] \cite{levy1979, lieb1983}
\be\label{FSearch}
F[n]\equiv\min_{\Psi \to n}\langle\Psi|\hat T+\hat U|\Psi\rangle
\ee
where $|\Psi\rangle$ is an $N$-particle state with wave function $\Psi$ of the correct symmetry (completely antisymmetric for fermions), $\hat T$ is the kinetic energy operator, $\hat U$ is the two-body interaction energy operator. The minimization is carried over the set of wave functions that yield the prescribed density $n(\rv)$ (positive, integrating to the total particle number $N$)\footnote{This definition can be extended to allow the search to run over ensembles of states that yield the prescribed density.  Such an extension is, of course, mandatory in the case of fractional particle numbers.} The appeal of this definition is that $F[n]$ is {\it universal}, i.e.,  in principle, it needs to be calculated only once and applies to all electronic systems, atoms, molecules, or solids.  Different systems differ only by the external potential $V(\rv)$: their ground state energy, $E_0$, is obtained by solving the minimization problem
\be
E_0=\min_{n}\left\{F[n]+\int n(\rv)V(\rv)d\rv\right\}\,.
\ee
The $F$ functional can further be separated into three components
\be
F[n]=T_s[n]+E_H[n]+E_{xc}[n]\,,
\ee
where $T_s[n]$ is the {\it non-interacting} kinetic energy defined as
\be\label{TSearch}
T_s[n]\equiv\min_{\Psi \to n}\langle\Psi|\hat T|\Psi\rangle\,,
\ee
$E_H[n]=\frac{1}{2}\int d\rv\int d\rv' n(\rv)U(\rv,\rv')n(\rv')$ is the classical potential of the particle distribution, (excluding correlation and including an unphysical self-interaction of each particle with itself) $E_{xc}[n]$ is the remainder.

 In the Kohn-Sham theory, the functional $T_s[n]$ does not appear explicitly: it is implicitly determined as the sum of the kinetic energies associated with the orbitals of the self-consistent solution of Kohn-Sham equations: hence most efforts focused on approximating $E_{xc}$ and $V_{xc}$. There are, however, some drawbacks to the Kohn-Sham procedure.  First of all, the solution of the self-consistent field (SCF) equation is relatively slow (computational time scaling as $N^3$).  Second, not knowing $T_s[n]$ as a functional of density precludes us from extracting $E_{xc}[n]$ for benchmarking purposes from accurate quantum mechanical solutions of many-body systems (e.g., configuration interaction \cite{leininger1997, sherrill1999}, quantum Monte Carlo \cite{hammond1994, nightingale1998, foulkes2001, santos2003}).  A more basic reason for trying to explicitly compute $T_s[n]$ and its functional derivative ($\frac{\delta T_s[n]}{\delta n(\rv)}=-V_s[n](\rv)$, also known as the Kohn-Sham potential) emerges in the context of Machine Learning (ML) and Artificial Intelligence (AI) as applied to electronic structure and materials science \cite{sun2024, zhong2024}.  It appears that the $T_s$ functional could be ``learned" by a neural network, given the relatively simple nature of the basic variable $n(\rv)$ and the relative ease of producing large amounts of training data by simply solving the Kohn-Sham equation -- an operation that is routinely performed thousands of times every day. \cite{snyder2012, seino2018, hollingsworth2018, meyer2020, imoto2021, ryczko2022, mazo2023, gangwar2023, sun2024, martinetto2024}

 At first sight, the computation of $T_s[n]$ from the density according to Eq.~(\ref{TSearch}) does not appear to be prohibitively difficult.  Indeed, the problem has been attacked many times in the past \cite{holas1991, wu2003, snyder2012, shi2021, shi2022n2v, gould2014, gould2023, mi2023} and some approximate treatments have been devised \cite{finzel2016, gangwar2023, sun2024}. This functional, for fermions,  separates  into two parts:
 \be\label{TB&TP}
 T_s[n]=T_B[n]+T_P[n]\,,
 \ee
 where the first term
 \be\label{TBosonic}
 T_B[n]=\int \frac{|\nabla n(\rv)|^2}{8 n(\rv)}d\rv\,,~~~~~~~(\hbar=m=1)
 \ee
 is the ``bosonic" functional, i.e., the kinetic energy of a system of non-interacting bosons in the ground state with density $n(\rv)$, and the second term is the ``Pauli" kinetic energy, i.e., the kinetic energy arising from the Pauli exclusion principle, which forces Fermions to occupy orthogonal orbitals. In the limit of large $N$, the Pauli functional is rigorously bound (from above) by the Thomas-Fermi functional \cite{thomas1927, fermi1927, lieb1981}, but this bound is too weak to be useful and gives no information about the functional derivative, i.e., the Pauli potential, which we denote by
 \be\label{VPauli}
 V_P[n](\rv)=\frac{\delta T_P[n]}{\delta n(\rv)}\,.
 \ee

 There are several algorithms for computing the Kohn-Sham potential from the density (a process known as the inversion of the Kohn-Sham equation), and from these algorithms, $T_s[n]$ can also be extracted\cite{shi2022n2v}. However, none of these algorithms is fast and accurate enough to provide the massive data input that is required to train modern AI networks.  Furthermore, computing $T_s[n]$ alone is not sufficient: one also needs its functional derivative $V_s[n]$. Recently, a ridge regression method was used to fit $T_s[n]$ for simple one-dimensional systems \cite{snyder2012},  but the calculation of the functional derivative remained problematic.

 Knowledge of the Pauli potential as a functional of the density would open spectacular vistas in DFT. It would enable us to solve material science problems in terms of density alone, without relying on Kohn-Sham orbitals.   In this orbital-free scheme, for example, the ground state density of a non-interacting system could be found directly from the solution of the bosonic problem
 \be
 \left\{-\frac{1}{2}\nabla^2 +V_P[n](\rv)+V(\rv)-\mu\right\}\sqrt{n(\rv)}=0\,.
 \ee

 In this paper, we propose a new method to calculate $T_P[n]$ and $V_P[n](\rv)$ by direct solution of the constrained minimization problem~(\ref{TSearch}).  We focus on a class of densities that has received little attention in the past, namely periodic densities.
Besides being everywhere positive and continuous, a periodic density satisfies periodic boundary conditions over a region of space of volume $\Omega$, which we call {\it the unit cell}, but not over any smaller region.
The complete density is constructed by assembling several identical unit cells, their number being denoted by $N_\Omega$.  The union of all these unit cells will be referred to as the {\it supercell}. The total number of electrons in the supercell is required to be an integer $N$, but the number of electrons per unit cell, i.e., $\nu=N/N_\Omega$, can be fractional~\footnote{We note that in a perfect crystal, the number of electrons per unit cell is an integer, because there is an integer number of atoms in each unit cell. However fractional electron numbers are possible when we consider defects, extrinsic doping, or even simply an effective periodic potential that does not arise from an assembly of neutral atoms.}.    While the density is, by definition, periodic over the unit cell,  the wave function does not necessarily have this property:  it can acquire a phase factor $e^{i\kv \cdot\Rv}$ as all the electron coordinates are translated by a lattice vector $\Rv$, connecting one unit cell to the next.  However, we still require the wave function to satisfy periodic boundary conditions over the supercell.  This restricts the set of admissible values of the wave vector $\kv$ to the usual Born-von Karman quantized values in the first Brillouin zone \cite{ashcroft1976}.
In the infinite periodic case, the number of unit cells tends to infinity together with the number of electrons, in such a way that the number of electrons per unit cell remains constant. At the same time, the wave vectors $\kv$ become uniformly distributed over the first Brillouin zone, the density of the distribution being $N_\Omega\Omega/(2\pi)^d$, where $d$ is the number of spatial dimensions.

 To perform the minimization of the energy at constant density $n(\rv)$ we introduce a basis of {\it equidensity orbitals} (EO)  \cite{harriman1981, zumbach1983, zumbach1984, ghosh1985, harriman1985, morrison1990, gygi2023, desantis1998} in the Hilbert space of one-electron wave functions.  This is a complete set of orthogonal wave functions, which all have the same density and satisfy periodic boundary conditions on the supercell.  They can be combined to form a complete set of orthogonal antisymmetric $N$-particle wave functions, which all have the desired density $n(\rv)$. A linear superposition of these wave functions does not, in general, have the same density $n(\rv)$, due to interference between different terms.  Nevertheless, the condition that a linear superposition of EOs preserves the density can be expressed as a simple condition on the coefficients of the superposition. This will be derived in Section~\ref{SectionIII}.
 {\it Crucially, this condition does not explicitly depend on the density: information about the density is entirely built into the choice of the EO basis.}  This ``universality" of the fixed density constraint is the principal technical advantage of our search method and, as we shall see, is the reason why we shall be able to calculate the functional derivative of $T_s[n]$ symbolically, after $T_s[n]$  is obtained.

 This paper is organized as follows: In Section~\ref{SectionII}, we introduce the equidensity orbitals, an adaptive basis set used in our calculations. In Section~\ref{SectionIII}, we describe the calculation of the kinetic energy functional for a system consisting of a single unit cell. In Section~\ref{ManyCells}, we extend the calculation to an arbitrary number of unit cells. In Section~\ref{constraint_impl}, we describe the implementation of the orthogonality and fixed-density constraints. In Section~\ref{inft_periodicity_scaling}, we describe the scaling properties of the kinetic energy functional. In Section~\ref{functional_derivative}, we derive a symbolic expression for the functional derivative of the kinetic energy functional, presenting a closed-form expression for the Pauli potential in 1-D. Subsequently. In Section~\ref{derivative_disc}, we address the derivative discontinuity in the Pauli potential as given by the HOMO-LUMO gap.
 Section~\ref{occupation_mask} addresses the determination of the occupied states in one dimension. Finally, in Section\ref{benchmarking}, we assess the performance of our algorithm versus exactly solved prototype systems of non-interacting spinless fermions in smooth 1-D periodic potentials. Section~\ref{outlook} summarizes our work and proposes directions of future research.

 \section{Equidensity orbitals}
 \label{SectionII}
 We consider a periodic system with a cubic unit cell of size $a$ and volume $\Omega=a^3$.  The supercell, denoted by ${\cal B}$,  contains $N_\Omega=L^3$ unit cells.
 The density is periodic over the unit cell, i.e.,
 \be
 n(\rv+\Rv)=n(\rv)
 \ee
 where $\Rv = a(n_1 \xv+n_2\yv+n_3\zv)$ is a lattice vector for  integers $n_1,n_2,n_3$.  The wave functions are periodic over the supercell with superlattice vectors $L\Rv$.

 Following Harriman, we introduce the {\it equidensity orbitals} (EO)
 \be\label{EO}
 \phi_\kv(\rv)\equiv\sqrt{\frac{n(\rv)}{N}}e^{i \kv \cdot \xiv(\rv)}\,,
 \ee
 where $N=\int_{\cal B} d\rv n(\rv)$ is the total number of electrons\footnote{Here and in the following we ignore spin.} and $\xiv: {\cal B}\to {\cal B} $ is an invertible map of the supercell into itself, such that its Jacobian determinant is
 \be\label{JacobianProblem1}
\det \left(\frac{\partial \xi_i(\rv)}{\partial r_j}\right) = \frac{n(\rv)}{N}
 \ee
 and
 \be\label{JacobianProblem2}
 \xi(\rv+\Rv)= \xi(\rv)+\Rv\,.\footnote{This implies that $\xiv(\rv)$ can be expressed as $\rv+u(\rv)$ where $u: \Omega\to \Omega$ is an invertible map of the unit cell into itself.}
 \ee
 The wave vectors $\kv$ are chosen so that the orbitals satisfy Born-von Karman periodic boundary conditions in the supercell, i.e.,
 \be\label{BVK}
 \kv = \frac{2 \pi}{La}(n_1 \xv+n_2\yv+n_3\zv)\,,
 \ee
 with $n_1,n_2,n_3$ integers.
 With these definitions, it is immediately clear that the equidensity orbitals form a complete set of orthonormal orbitals in the Hilbert space of one-particle wave functions that are periodic over ${\cal B}$.  This is because
 \ber
 \int_{\cal B} d\rv \phi^*_\kv(\rv)\phi_{\kv'}(\rv) &=& \int_{\cal B}d\rv \frac{n(\rv) }{N}e^{-i(\kv-\kv')\cdot\xiv(\rv)}\nn\\
 &=& \int_{\cal B}d\xiv e^{-i(\kv-\kv')\cdot\xiv(\rv)}=\delta_{\kv,\kv'}\,.
 \eer
 Furthermore, each orbital has a density $n(\rv)/N$ associated with it. Therefore, if we choose a set of $N$ distinct wave vectors $\{\kv_i\}\equiv \{\kv_1,\kv_2,...,\kv_N\}$, then the Slater determinant constructed from the corresponding orbitals and denoted by ${\cal S} (\{\kv_i\})=[N!]^{-1/2}\det \left[\phi_{\kv_i}(\rv_j)\right]$, with $\rv_1,\rv_2,...,\rv_N$ the electronic coordinates and $i,j$ taking values $1-N$, will have a density $n(\rv)$.  The most general $N$-particle antisymmetric wave function, regardless of its density, is expressible as a linear combination of such Slater determinants.

 Notice that we can ignore the spin as long as the spin density is collinear: the total density breaks up into spin-up and spin-down contributions, $n(\rv)=n_\up(\rv)+n_\down(\rv)$, and the noninteracting kinetic energy functional is simply the sum of two spinless functionals: $T_s[n_\up,n_\down]=T_s[n_\up]+T_s[n_\down]$.

 \section{Construction of the kinetic energy functional for a single periodic cell}\label{SectionIII}
 In this section, we assume that the supercell and the unit cell coincide, i.e., we set $N_\Omega=1$.
 The expectation value of the kinetic energy in a  many-body quantum state described by a wave function $\Psi(\rv_1, \rv_2, ..., \rv_N)$ (normalized to $1$) is most conveniently expressed in terms of the reduced one-body density matrix
 \be\label{GammaMatrix}
 \gamma (\rv,\rv') \equiv N\int d\rv_2...d\rv_N \Psi^*(\rv,\rv_2,...,\rv_N)\Psi(\rv',\rv_2,...,\rv_N)\,,
 \ee
 and is given by
 \be
 \langle \Psi|\hat T|\Psi\rangle = \frac{1}{2}\int d\rv  \left.\nablabold_{\rv} \cdot \nablabold_{\rv'} \gamma(\rv,\rv')\right\vert_{\rv'=\rv}\,.
 \ee

 We now introduce the equidensity orbital representation of the density matrix
 \be\label{GammaEO}
 \gamma(\rv,\rv')=\sum_{\Gv,\Gv'}\gamma_{\Gv,\Gv'} \phi_{\Gv}(\rv)\phi^*_{\Gv'}(\rv')\,.
 \ee
 where $\Gv$,$\Gv'$ are wave vectors of the form
\be\label{RLV}
 \Gv = \frac{2 \pi}{a}(n_1 \xv+n_2\yv+n_3\zv)\,,
 \ee
 such that the density matrix is periodic over the (single) unit cell.  Then, making use of Eqs.~(\ref{GammaEO}) and (\ref{EO}), it is straightforward to see  that the kinetic energy is the sum of three terms:

\be\label{Tdecomposition1}
T^{(1)}=\frac{1}{2N}\int d\rv \left\vert \nablabold \sqrt {n(\rv)}\right\vert^2 \sum_{\Gv,\Gv'} \gamma_{\Gv,\Gv'} e^{-i(\Gv'-\Gv)\cdot\xiv(\rv)}\,,
\ee
\be\label{Tdecomposition2}
T^{(2)}= \frac{1}{2N}\int d\rv \nablabold  n(\rv) \cdot \sum_{\Gv,\Gv'} \gamma_{\Gv,\Gv'} e^{-i(\Gv'-\Gv)\cdot\xiv(\rv)} \nablabold_{\rv}[(\Gv-\Gv')\cdot\xiv(\rv)] \,,
\ee
\be\label{Tdecomposition3}
T^{(3)}= \frac{1}{2N}\int d\rv ~ n(\rv) \sum_{\Gv,\Gv'} \gamma_{\Gv,\Gv'} e^{-i(\Gv'-\Gv)\cdot\xiv(\rv)} \nablabold_{\rv}[\Gv\cdot\xiv(\rv)]\cdot \nablabold_{\rv}[\Gv'\cdot\xiv(\rv)] \,.
\ee
The above formulas are valid for any state.  Now let us include the additional hypothesis that the electronic density of the state is $n(\rv)$, {\it i.e., it is the same as the density that was used to define the EO orbitals}.   This means that $\gamma(\rv,\rv)=n(\rv)$.  Substituting the  expressions for $\gamma$ and $\phi_{\kv}$ we get
\be
n(\rv)=\sqrt{\frac{n(\rv)}{N}} \sum_{\Qv}\sqrt{\frac{n(\rv)}{N}} e^{-i\Qv\cdot\xiv(\rv)} \sum_{\Gv}\gamma_{\Gv,\Gv+\Qv}\,,
\ee
which, because of the orthogonality and completeness of the EOs, implies
\be\label{SubtraceConstraint}
\sum_{\Gv}\gamma_{\Gv,\Gv+\Qv}=N \delta_{\Qv,0}\,,
\ee
for any  vector $\Qv$ of the form~\ref{RLV}.
In other words, the trace of the one-particle density matrix is $N$, and all the ``subtraces", i.e., the sums of matrix elements along lines parallel to the diagonal, are zero.  We note that this constraint on the form of the one-particle density matrix does not explicitly depend on the density.

Making use of Eq.~(\ref{SubtraceConstraint}) in Eqs.~(\ref{Tdecomposition1}--\ref{Tdecomposition3})  we find
\be
T^{(1)}=T_B[n]\,,
\ee
where $T_B[n]$ is the bosonic functional defined in Eq.~(\ref{TBosonic}),  and also \be
T^{(2)}= 0
\ee
and
\be
T^{(3)}=  \frac{1}{8N}\int d\rv ~ n(\rv) \sum_{\Gv,\Gv'} \gamma_{\Gv,\Gv'}  e^{-i(\Gv'-\Gv)\cdot\xiv(\rv)} \left\vert\nablabold_{\rv}[(\Gv+\Gv')\cdot\xiv(\rv)]\right\vert^2 \,.
\ee
The $T^{(3)}$ term will give the Pauli functional after being minimized with respect to all the  ``$n$-representable" one-particle density matrices (i.e., the density matrices that can arise from an antisymmetric many-body wave function according to Eq.~\ref{GammaMatrix}) that also satisfy the constraint~(\ref{SubtraceConstraint}).  In principle, it is not necessary to minimize with respect to $\xiv(\rv)$, provided it satisfies the conditions stated in Eqs.~(\ref{JacobianProblem1}--\ref{JacobianProblem2}). In practice, because the calculation is necessarily done on a finite subset of EO's, the result of the minimization depends on the choice of $\xiv$, and should be optimized with respect to the latter.

\begin{figure}[t]
    \includegraphics[width=0.60\textwidth]{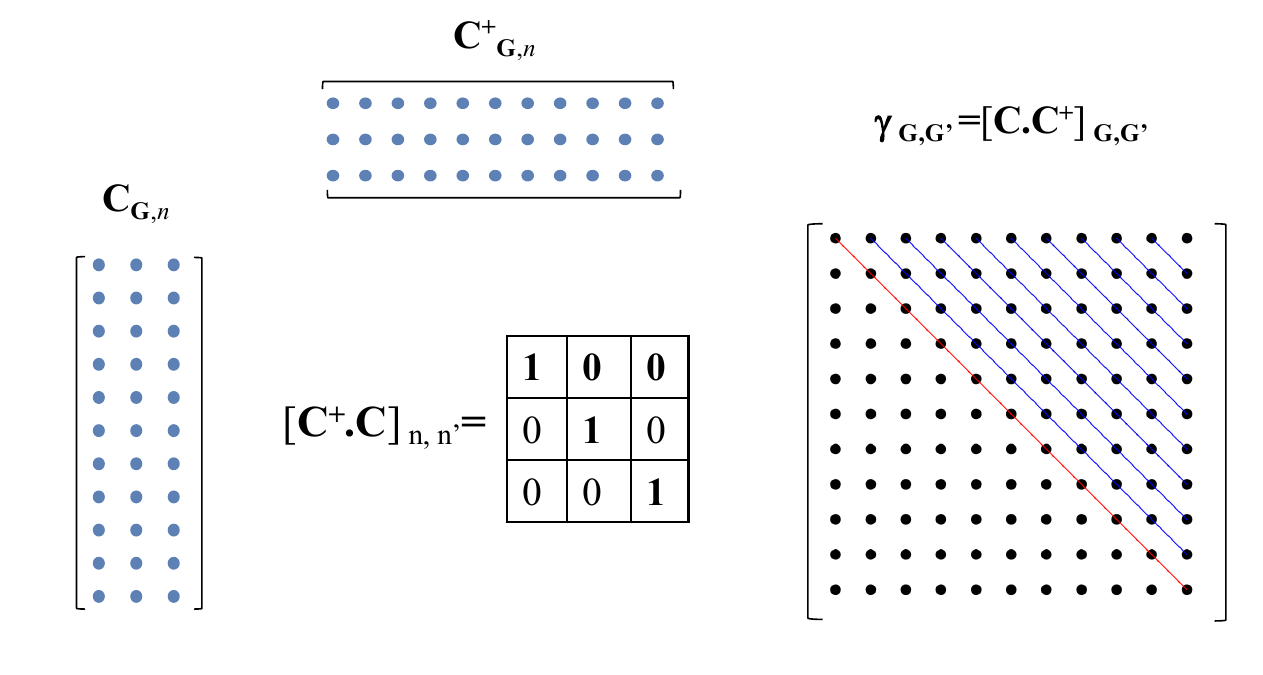}
    \caption{Schematic representation of the matrices of orbital coefficients $\Cv$, $\Cv^\dagger$, the density matrix $\gammav$ and its trace (red line) and subtraces (blue lines), whose vanishing ensures the density constraint.}
    \label{fig:Matrices}
\end{figure}

The condition for $n$-representability of $\gamma$ is well known.  It is necessary and sufficient that all its eigenvalues be comprised between $0$ and $1$.  This constraint, together with Eq.~~(\ref{SubtraceConstraint}), completely defines the search.  However, from physical considerations, we expect that the solution to the minimization problem for the kinetic energy should be a density matrix associated with a single Slater determinant or, at most, with an ensemble of single Slater determinants.  The density matrix associated with a single Slater determinant satisfies the {\it idempotency condition} $\gamma^2=\gamma$, which implies that all its eigenvalues are either $0$ or $1$.
One way to ensure the idempotency condition is to represent $\gamma$ -- an $M\times M$ hermitian matrix of the form
\be
\gamma= \Cv \cdot \Cv^\dagger \,,
\ee
where $\Cv$ is a rectangular $M \times N$ matrix with $M\gg N$ corresponding to the number of basis states, $\Cv^\dagger$ is its ``Hermitian conjugate" (a rectangular $N\times M$ matrix obtained by transposing and complex-conjugating $\Cv$) and furthermore
\be\label{NormalizationCondition}
 \Cv^\dagger \cdot \Cv = {\bf 1}_{N\times N}
 \ee
where ${\bf 1}_{N\times N}$  is the $N\times N$ identity matrix. A schematic representation of these matrices is given in Fig.~\ref{fig:Matrices}. It is evident from these formulas that $\gammav$ is idempotent and its trace is $N$.\footnote{The astute reader will realize that the rows of $C$, denoted by $C_{i,\kv}$, define $N$ orthonormal orbitals $\psi_i(\rv)=\sum_\kv C_{i,\kv}\phi_{\kv}(\rv)$, with Eq.~(\ref{NormalizationCondition}) ensuring orthonormality.  Thus, our ``orbital-free" formulation of DFT is not so orbital-free after all!}

Denoting the Pauli kinetic energy of a given EO $\Cv$ as
 \be\label{TP}
  T_P[\Cv, n] =  \sum_{\Gv,\Gv'}[\Cv \cdot \Cv^\dagger]_{\Gv,\Gv'}T_{\Gv',\Gv}[n]\,,
 \ee
  Our problem can be formulated as follows: For a given non-negative periodic $N$-electron density $n(\rv)$, the Pauli part of the kinetic energy functional will be given by
 \be\label{ConstrainedSearch}
  T_P[n] = \min_{\Cv \in {\cal M}} T_P[\Cv, n]
 \ee
 where
 \be \label{TK}
 T_{\Gv,\Gv'}[n] \equiv \frac{1}{2N}\int d\rv ~ n(\rv) e^{-i(\Gv'-\Gv)\cdot\xiv(\rv)} \left\vert\nablabold_{\rv}[(\Gv+\Gv')\cdot\xiv(\rv)] \right\vert^2
\ee
depends on $n(\rv)$ both explicitly and implicitly, i.e., through $\xiv(\rv)$.
The manifold ${\cal M}$ on which the minimum must be sought is defined by the conditions
\be\label{C-Manifold}
  \Cv^\dagger \cdot \Cv = {\bf}1_{N\times N}\,,~~~~ {\rm STr_q}[ \Cv\cdot \Cv^\dagger]=0~~~{\rm for~} 1\leq q \leq M-1\,,
\ee
where ${\rm STr}_q$ denotes the $q$-th subtrace of a matrix, i.e., the sum of the matrix elements on the $q$-th ``subdiagonal", where the label $q$ starts at $q=1$ for the subdiagonal immediately on the right of the true diagonal and ends at $M-1$ for the last subdiagonal consisting of a single corner element of $\gammav$ (see Fig.~\ref{fig:Matrices}).

Denoting by $\bar \Cv[n]$ the minimizer of Eq.~(\ref{ConstrainedSearch}) our final expression for the Pauli energy functional becomes
\be\label{PauliFunctional}
 T_P[n] =  \sum_{\Gv,\Gv'}\left\{\bar \Cv[n] \cdot \bar \Cv^\dagger[n]\right\}_{\Gv,\Gv'}T_{\Gv',\Gv}[n]\,.
\ee

 \section{Extension to an arbitrary number of unit cells}\label{ManyCells}
 The formulas of the previous section were derived for the case of a single unit cell, $N_\Omega=1$, which coincides with the supercell. Thus, the periodicity of the density coincided with the periodicity of the wave functions. Now, we extend the formulation to consider a supercell that contains an integer number of unit cells, $N_\Omega>1$, so that the periodicity of the wave function (across the supercell) and that of the density (across the unit cell) differ.  According to Bloch's theorem,  the single-particle states are labeled by a band index $n$ and a Bloch wave vector $\kv$ of the form~\ref{BVK} belonging to the first Brillouin zone, i.e. the maximal set of distinct wave vectors $\kv$ (quantized according to the size of the supercell, as in Eq.~(\ref{BVK})) that cannot be connected by a reciprocal lattice vector $\Gv$ (quantized according to the size of the unit cell, as in Eq.~(\ref{RLV}).  Now the rectangular matrix $\Cv$ becomes a function of $\kv$,
 whose  dimension is $M \times N_\kv$ where $M$ is, as in the previous section, the number of basis functions of the form
 \be\label{EO}
 \phi_{\kv+\Gv}(\rv)\equiv\sqrt{\frac{n(\rv)}{N}}e^{i (\kv+\Gv) \cdot \xiv(\rv)}\,,
 \ee
 which we use to expand the Bloch wave function, and $N_{\kv}$ is the number of occupied bands at wave vector $\kv$.  The total number of electrons is given by
 \be
 N=\sum_{\kv \in {\rm BZ}}N_{\kv}=\tilde N N_\Omega\,,
 \ee
  $\tilde N$ is the number of electrons per unit cell. Since there are exactly $N_\Omega$ wave vectors in the first Brillouin zone, we see that $\tilde N$ is the average of $N_{\kv}$ over the  Brillouin zone\footnote{In general, the distribution of the occupations $N_{\kv}$ over the  Brillouin zone must be determined by energy minimization.  In one dimension the solution of this ``occupation problem''  is suggested by known exact features of the band structure, but more complicated cases may arise in higher dimensions as different energy bands overlap.}.  With  this being said, the constrained minimization problem for a cell density $n(\rv)$ and $N_\Omega$ unit cells takes the form
   \be\label{TPManyCells}
T_P[n,N_\Omega]=\min_{\Cv \in {\cal M}} T_P[\Cv, \kv, n,N_\Omega]\,,
 \ee
  where
\be
T_P[\Cv,  n,N_\Omega]= \sum_{\kv \in {\rm BZ}}\sum_{\Gv,\Gv'}[\Cv(\kv) \cdot \Cv^\dagger(\kv)]_{\Gv,\Gv'}T_{\kv+\Gv',\kv+\Gv}[n]\,.
\ee
Here $T_{\kv+\Gv',\kv+\Gv}$ is still given by Eq.~(\ref{TK}) with $\Gv$ and $\Gv'$ replaced by $\kv+\Gv$ and $\kv+\Gv'$ respectively, the integral done over the unit cell and $N$ replaced by $\tilde N$.   The set of admissible $\Cv$ matrices, denoted by ${\cal M}$, is defined by
 \be\label{C-Manifold2}
 \Cv^\dagger(\kv) \cdot \Cv(\kv)={\bf 1}_{N_\kv\times N_\kv}\,,
 \ee
 a condition that automatically implies the idempotency of the $M\times M$ density matrix $\gamma(\kv)=\Cv (\kv) \cdot \Cv^\dagger(\kv)$ and the trace condition ${\rm Tr}\gamma(\kv)=N_\kv$.
 At the same time, the density constraint takes the form of a constraint on the subtraces of the density matrix:
 \be\label{Subtrace2}
g_{\Qv}(\Cv) = \sum_{\kv \in {\rm BZ}}\sum_{\Gv}[C(\kv) \cdot C^\dagger(\kv)]_{\Gv,\Gv+\Qv}=0\,,
 \ee
 where $\Qv \neq {\bf 0}$ is a non-null reciprocal lattice vector \footnote{Notice that the $\Qv=0$ component of the constraint, i.e., the condition that the trace of the density matrix equals $N$, is already contained in Eq.~(\ref{C-Manifold2}.}.  Notice that the orthogonality constraint~(\ref{C-Manifold2}) applies separately at each point $\kv$ in the Brillouin zone because states with different values of $\kv$ are automatically orthogonal.   However, the subtrace constraint~(\ref{Subtrace2}) includes a sum over $\kv$ and thus couples matrices at different wave vectors. This is the reason why the energy optimization problem does not split into independent optimizations at different $\kv$s.

\section{Implementation of constraints}\label{constraint_impl}

\begin{figure*}[ht]
    \centering
    \begin{subfigure}[b]{0.48\textwidth}
        \centering
        \includegraphics[width=\textwidth]{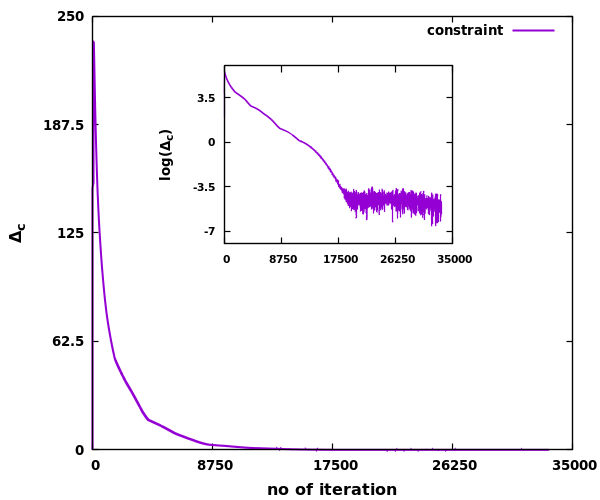}
        \caption{Constraint violation $\Delta_c$ as a function of iterations.}
        \label{fig:subfig1}
    \end{subfigure}
    \hfill
    \begin{subfigure}[b]{0.48\textwidth}
        \centering
        \includegraphics[width=\textwidth]{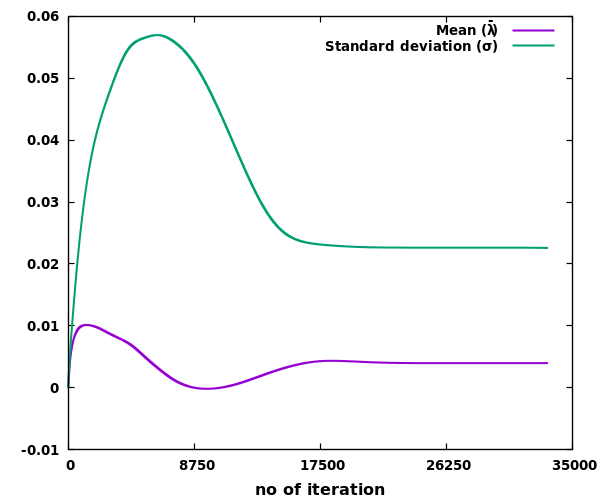}
        \caption{Evolution of mean Lagrange multiplier $\bar{\lambda}$ and standard deviation $\sigma$.}
        \label{fig:subfig2}
    \end{subfigure}

    \caption{(a) Behavior of constraint violation, quantified as $\Delta_c=G^{-1}\left(\sum_{\Qv\neq 0}\left \vert\sum_{\kv \in {\rm BZ}}\sum_{\Gv}[C(\kv) \cdot C^\dagger(\kv)]_{\Gv,\Gv+\Qv}\right \vert \right)$, over iterations. Inset: Logarithm of $\Delta_c$.
    (b) Evolution of the Lagrange multiplier mean value, $\bar{\lambda} = \frac{1}{G} \sum_{\Qv = 0}^{\Gv} \lambda_{\Qv}$, and standard deviation, $\sigma = \sqrt{\frac{1}{G}\sum_{\Qv=0}^{\Gv} (\lambda_\Qv - \bar{\lambda})^2}$, during optimization. Parameters: $N=22$, $N_{\Omega}=15$.}
    \label{fig:constraint_convg}
\end{figure*}

The two constraints expressed in Eq.~(\ref{C-Manifold2})  and ~(\ref {Subtrace2})  are not easily satisfied simultaneously.  In our implementation of the minimization algorithm, the orthogonality constraint~(\ref{C-Manifold2}) is enforced through the QR decomposition \cite{trefethen2022numerical}. Starting from an arbitrary $M \times N$ matrix $\Xv$ the QR decomposition factors it as the product of an orthogonal $M \times N$ matrix $Q(\Xv)$ and an upper triangular $N\times N$ matrix $R(\Xv)$ where $\Xv=Q(\Xv)R(\Xv)$. Setting $\Cv$ as the orthogonal part
\be
\Cv = Q(\Xv)
\ee
ensures the orthonormality of the columns of $\Cv$ and the rows of $\Cv^\dagger$. With this, we can reparameterize the Pauli kinetic energy and the density constraint in terms of the unconstrained parameter $\Xv$:
\be
\bar{T}_P[\Xv, n, N_{\Omega}]:=T_P[Q(\Xv), n, N_{\Omega}], \quad \bar{g}_\Qv(\Xv) =  g_{\Qv}(Q(\Xv)).
\ee

Unlike the orthonormality constraint, the density constraint (\ref{Subtrace2}) is hard to reparameterize. We resort to the Lagrangian multiplier method. The Lagrangian function to be minimized  is given as
\be
  \mathcal{L}[\Xv, \lambda_{\Qv} ] = \bar{T}_P[\Xv, n, N_{\Omega}] + \sum_{\Qv\neq 0}\lambda_{\Qv} \bar{g}_{\Qv}(\Xv)
\ee
whose stationary conditions
\begin{equation}
  \left\{\begin{array}{l}
 \pdv{\mathcal{L}}{\Xv} = -\pdv{\overline{T}_{p}}{X_{i}} - \sum_{\Qv} \lambda_{\Qv} \pdv{\overline{g}_{\Qv}}{X_{i}} = 0    \\
 \pdv{\mathcal{L}}{\lambda _{\Qv}} =-\overline{g}_{\Qv}(\vb{X}) = 0    \\
  \end{array}\right.
\end{equation}
give the stationary point of $\bar{T}_{P}$ where the constraint $\pdv{\mathcal{L}}{\lambda_{\Qv} }=\bar{g}_{\Qv}(\Xv)=0$ is satisfied, and, in the process,  determine the values of the Lagrange multipliers.
To solve these equations numerically, we let $X_i$ and $\lambda_\Qv$ evolve under the fictitious dynamics
\begin{equation}
\begin{split}
  \dot{X_{i}} =&  -\pdv{\overline{T}_{p}}{X_{i}} - \sum_{\Qv} \lambda_{\Qv} \pdv{\overline{g}_{\Qv}}{X_{i}}, \\
\dot{\lambda}_{\Qv} =& \overline{g}_{\Qv}(\vb{X})\,,
\end{split}
\end{equation}
which reaches equilibrium at the stationary points.
Taking a second derivative and substituting $\dot{\lambda}_{\Qv}$ with $\overline{g}_{\Qv}(\vb{X})$ leads to a closed equation of motion for $\Xv$, which is second order in the fictitious time and resembles the equation of motion of a damped harmonic oscillator
\begin{equation}
  \ddot{X}_{i} + \sum_j \left( \underbrace{\pdv[2]{\overline{T}_{P}}{X_{i}}{X_{j}} + \sum_{\Qv} \lambda_{\Qv} \pdv[2]{\overline{g}_{\Qv}}{X_{i}}{X_{j}}}_{A_{ij}} \right) \dot{X}_{j} + \sum_{\Qv} \overline{g}_{\Qv} \pdv{\overline{g}_{\Qv}}{X_{i}} = 0
\end{equation}
with the damping matrix $A_{ij}$ and the potential energy $U(\Xv)=\sum_{\Qv} \frac{1}{2}\overline{g}_{\Qv}(\Xv)^{2}$. The time derivative of the total energy, kinetic plus potential, is
\begin{equation}
  \dot{E} = \dv{}{t} \left( \sum_i \frac{1}{2} \dot{X}_{i}^{2}  \right) + \dot{U}
  = - \sum_{ij} \dot{X}_{i} A_{ij} \dot{X}_{j}\,.
\end{equation}
We see that a positive definite $A_{ij}$ causes a monotonic decrease of the energy, driving the system towards equilibrium, which is also the solution to the Lagrangian problem.

There is one problem left however: the $A_{ij}$ is purely determined by the form of $\overline{T}_{P}$ and $\overline{g}_{\Qv}$ and may not be positive definite.
To tackle this, we employ the Modified Differential Multiplier Method (MDMM) \cite{platt1987}. Specifically, we replace $\lambda _{\Qv}$ with $\lambda _{\Qv} + c_{\Qv} \overline{g}_{\Qv}(\Xv)$ in $\dot{X} _{i}$, which leads to a new damping matrix $A'_{ij}$:

\begin{equation}\label{damping_coeff}
  A'_{ij} =  A_{ij} + \sum_{\Qv}  c_{\Qv}\pdv{\overline{g}_{\Qv}}{X_{i}}\pdv{\overline{g}_{\Qv}}{X_{j}} + c_{\Qv} \overline{g}_{\Qv} \pdv{\overline{g}_{\Qv}}{X_{i}}{X_{j}}
\end{equation}

without changing the equilibrium solution.
$A'_{ij}$ is positive definite for large enough damping coefficients $c_{\Qv}$ according to a theorem in \cite{platt1987}.  We find that we can ensure positivity by setting  $c_{\Qv}$ to $10$ for all $\Qv$ in our experiments, regardless of the system. Furthermore, we used Adam \cite{kinga2015} optimizer to compute all the gradients, which adds momentum and estimated second-order information for better convergence.

The quality of constraint satisfaction, illustrated in Fig.~\ref{fig:constraint_convg}, is excellent, with the sum of the subtraces becoming as small as $10^{-4}$  with Lagrange multiplier $\lambda$ on the order of $10^{-2}$.

\section{Infinite periodicity and scaling}\label{inft_periodicity_scaling}
Because periodic systems can be extended indefinitely by adding more and more unit cells it is important to establish how the functional $T_s[n,N_\Omega]$ behaves as the number of unit cells, $N_\Omega$ tends to infinity while the density $n(\rv)$ in each unit cell remains fixed.  Because of the extensivity of the kinetic energy we expect that in the limit of large $N_\Omega$,
\be\label{Extensivity}
 T_s[n,N_\Omega]\stackrel{N_\Omega\to \infty} \to N_\Omega \bar T_s[n]\,,
 \ee
  where, on the left,  the functional $\bar T_s[n]$ depends only $n(\rv)$\footnote{The extensivity relation~(\ref{Extensivity}) is not expected to hold for finite $N_\Omega$.  For example, for $N_\Omega=2$, the kinetic energy associated with the density in one unit cell will be affected by the change in boundary conditions on the wave function caused by the presence of the second unit cell (i.e., the wave functions in each unit cell can be either periodic or antiperiodic over the unit cell).  The dependence of the energy on boundary conditions is expected to disappear for $N_\Omega \to \infty$ as the wave function can acquire any phase between $0$ and $2\pi$ across the unit cell.}.

\begin{figure}[htb]  
    \centering
    \includegraphics[width=0.50\textwidth]{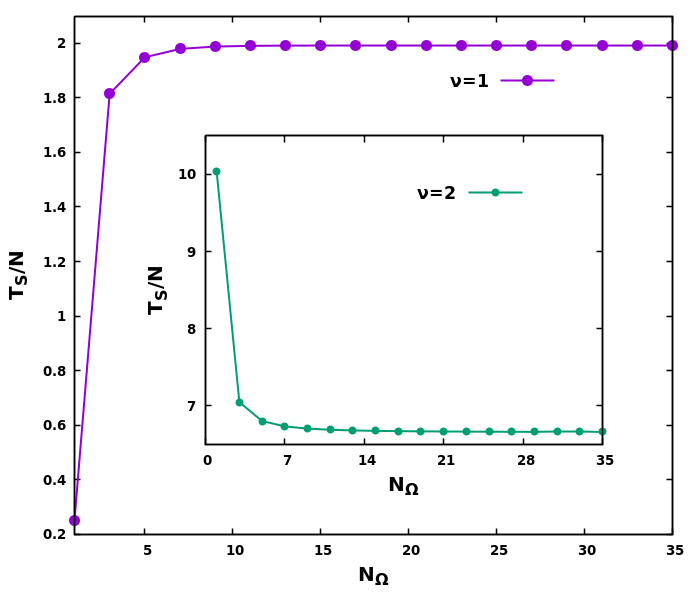}  
    \caption{Behavior of $\frac{T_{s}}{N}$, in units of $\frac{\hbar^2}{ma^2}$ as a function of N$_{\Omega}$ for filling factors $\nu \equiv N/N_\Omega=1$ and $\nu=2$ (inset). The infinite periodicity limit is reached for $N_\Omega=10$ in the first case and $N_\Omega=15$ in the second.}
    \label{fig:fig_ts_convg}
\end{figure}

This raises the interesting question of whether it is possible to directly calculate the intensive functional $\bar T_s[n_\Omega]$ without doing a supercell calculation. In the limit of infinite $N_\Omega$ the formulation presented in Section~\ref{ManyCells} remains valid, with the understanding that the quantized wave vectors in the first Brillouin zone are replaced by a continuous variable and the sum over discrete wave vectors is replaced by an integral over the first Brillouin zone.  Apart from this change, the key equations (\ref{TPManyCells}) and (\ref{C-Manifold2}) remain valid.

In practice, the integral over the Brillouin zone can be approximated, as accurately as desired, by a sum over a discrete grid of $\kv$ points, which does not have to change as $N_\Omega\to\infty$. This is because the subspace of occupied states is generally a smooth function of $\kv$  (for the handling of  Fermi surface discontinuities in metallic systems see Section~\ref{occupation_mask}).  Therefore, the limit of Eq.~(\ref{Extensivity}) is reached rapidly and $\bar T_s[n]$ is obtained without difficulty.  In our calculations, we find that $\bar N_\Omega=15$ is sufficient to reach the limit of infinite periodicity, as shown in Fig.~\ref{fig:fig_ts_convg}.

Next, we consider the exact scaling relation \cite{borgoo2014}
\be\label{ScalingRelation}
T_s[n_\lambda]=\lambda^2 T_s[n]\,,~~~n_\lambda(\rv)\equiv \lambda^d n(\lambda\rv)\,,
\ee
where $\lambda$ is a positive scale factor (no connection with the penalty method!) and $d$ is the dimensionality of the space.
When applied to a periodic density with lattice constant $a$ the scaling relation simply says that the natural units of the problem are $a$ for length (position), $a^{-d}$ for density, and $\hbar^2/(ma^2)$ for energy.  Constructing $T_s[n]$ in these units, as we will do in the rest of this paper, guarantees that any scaling transformation of the density, $\rv \to \lambda \rv$, $n \to \lambda^d n$ is absorbed in a rescaled lattice constant  $a \to a/\lambda$ and a rescaled unit of energy $\hbar^2 \lambda^2/(ma^2)$, in agreement with Eq.~(\ref{ScalingRelation}).

\section{Symbolic evaluation of the functional derivative}\label{functional_derivative}

We now address the problem of calculating the external potential $V_s(\rv)$ that produces the assigned density $n(\rv)$  in the ground state of $N$ non-interacting electrons.
First of all, it is evident that this potential can only be determined up to an arbitrary additive constant.  We resolve this ambiguity by demanding that $V_s(\rv)$ has zero average over the unit cell.
From the variational principle of DFT, we know that
\be
V_s(\rv)=-\left.\frac{\delta T_s[n]}{\delta n(\rv)}\right\vert_N
\ee
where the functional derivative is done at constant particle number $N$.
Splitting the kinetic energy into bosonic and Pauli components and making use of Eq.~(\ref{TBosonic}), one easily finds
\be\label{potcomp}
V_s(\rv)=\frac{\nabla^2 n^{1/2}(\rv)}{2 n^{1/2}(\rv)} -V_P(\rv)\,,
\ee
with $V_P(\rv)=\frac{\delta T_P[n]}{\delta n(\rv)}\big \vert_N$.  The main task is then the calculation of the functional derivative of the Pauli energy functional with respect to the density at constant particle number.
Remarkably, this can be done by essential symbolic manipulations (i.e., quadratures) as long as the solution of the optimization problem for the energy is available. At variance with all previous calculations of the Pauli potential, we do not require an explicit knowledge of the single-particle eigenvalues and their eigenfunctions.

The reason why this can be done is that the constrained space in which the variational parameters $\bf C$ are varied is defined by Eqs.~(\ref{C-Manifold}), which does not explicitly involve the density.  When the density is infinitesimally varied from $n(\rv)$ to $n(\rv) +\delta n(\rv)$ at constant particle number both the basis functions $\phi_{\kv+\Gv}(\rv)$ and the optimal coefficients  $\Bar {\bf C}[n]$  change infinitesimally and the first-order variation of $T_P[n]$ consists of two parts, one coming from the variation of  $\Bar {\bf C}[n]$ at constant basis, the other from the variation of the basis, which drives a variation of the matrix elements $T_{\kv +\Gv',\kv+\Gv}[n]$ at constant $\Bar {\bf C}[n]$.  However, $T_P[n]$ is stationary with respect to an infinitesimal variation of $\Bar {\bf C}[n]$ at a constant basis, as long as the variation respects the constraints imposed on the $\bf C$s, i.e., Eqs.~(\ref{C-Manifold}).  But this is precisely the case here because the constraint equations~(\ref{C-Manifold}) do not depend on density.   We conclude that the first order variation of $T_P[n]$ is entirely due to the variation of $T_{\kv+\Gv',\kv+\Gv}[n]$:
\be
\delta T_P[n]= \sum_{\kv \in {\rm BZ}}\sum_{\Gv,\Gv'}\left\{\bar {\bf C}[n] \cdot \bar {\bf C}^\dagger[n]\right\}_{\kv+\Gv,\kv+\Gv'}\delta T_{\kv+\Gv',\kv+\Gv}[n]\,,
\ee
while the optimal $\bf C$s can be treated as known constants.
This implies that the functional derivative of the Pauli energy functional  with respect to the density can be expressed in terms of the functional derivative of $T_{\kv+\Gv',\kv+\Gv}[n]$, which can be calculated symbolically if $\xiv(\rv)$ is known as an analytic functional of $n(\rv)$:
\be
\left.\frac{\delta T_P[n]}{\delta n(\rv)}\right\vert_N = \sum_{\kv \in {\rm BZ}}\sum_{\Gv,\Gv'}\bar \gamma_{\kv+\Gv',\kv+\Gv}[n] \left.\frac{\delta T_{\kv+\Gv',\kv+\Gv}[n]}{\delta n(\rv)}\right\vert_N\ ,
\ee
where;
\be
\bar \gamma_{\kv+\Gv',\kv+\Gv}[n]=\left\{\bar {\bf C}[n] \cdot \bar {\bf C}^\dagger[n]\right\}_{\kv+\Gv',\kv+\Gv}\,.
\ee
Making use of Eq.~(\ref{TK}) (where $\Gv$ and $\Gv'$ are replaced by $\kv+\Gv$ and $\kv+\Gv'$ respectively) we find
\ber \label{FunctionalDerivative}
&&\frac{\delta T_{\kv+\Gv',\kv+\Gv}[n]}{\delta n(\rv)}=\frac{1}{2N} e^{-i(\Gv'-\Gv)\cdot\xiv(\rv)}\left\vert\nablabold_{\rv}[(\kv+\Gv')\cdot\xiv(\rv)]\right\vert^2 \nn\\
&-&\frac{1}{N}\int d\rv' ~ \nabla_{\rv'}\left(n(\rv') e^{-i(\Gv'-\Gv)\cdot\xiv(\rv')}\right) \cdot \nabla_{\rv'}[(\kv+\Gv')\cdot\xiv(\rv')](\kv+\Gv')\cdot\frac{\delta \xiv(\rv')}{\delta n(\rv)}\nn\\
&-&\frac{1}{N}\int d\rv' ~ n(\rv') e^{-i(\Gv'-\Gv)\cdot\xiv(\rv')}\nabla_{\rv'}^2[(\kv+\Gv')\cdot\xiv(\rv')](\kv+\Gv')\cdot\frac{\delta \xiv(\rv')}{\delta n(\rv)}\nn\\
&-&\frac{1}{2N} \int d\rv' ~ n(\rv') e^{-i(\Gv'-\Gv)\cdot\xiv(\rv')}\left\vert\nablabold_{\rv'}[(\kv+\Gv')\cdot\xiv(\rv')]\right\vert^2 i (\Gv'-\Gv)\cdot\frac{\delta \xiv(\rv')}{\delta n(\rv)}\,.\nn\\
\eer

In the special case of one dimension, the map $\xi$ which transforms a uniform density $\bar n$ to a target density $n(x)$ takes the simple form
\be
\xi(x)=\int_0^x \frac{n(x')}{\bar n}dx'\,,
\ee
where $\bar n = a^{-1}\int_0^a n(x) dx$.
It is immediately verified that this is monotonically increasing, invertible, and satisfies the condition $\xi(x+a)=\xi(x)+a$ if $n(x+a)=n(x)$.  The functional derivative   $\delta\xi(x)/\delta n(x')=\bar n^{-1}\Theta(x-x')$ where $\Theta(x)$ is the Heaviside step function.

The construction of the map $\zeta$ which transforms a density $n_1(x)$ to a density $n_2(x)$ is equally straightforward. From Eq.~(\ref{JacobianProblem2}) we see that $\zeta$ is the composition of two maps, $\zeta=\xi_1^{-1}\circ\xi_2$, and can be explicitly constructed by solving the equation
\be
\int_0^{\zeta}\frac{n_1(y)}{\bar n}dy = \int_0^x \frac{n_2(y)}{\bar n}dy\,,
\ee
for $\zeta$ as a function of $x$.

For the Pauli potential in one dimension, assuming that the unit cell varies within $0 \leq x \leq a$, we have,
\begin{align}
V_P(x) &= \sum_{\kv,\Gv',\Gv} \bar \gamma_{\kv,\Gv',\Gv}[n]
    \frac{(\kv + \Gv)(\kv + \Gv')}{2 [\Bar{n}]^3}
    \Biggl[ 3 n(x)^2 e^{i(\Gv-\Gv')\cdot\xiv(x)} \notag \\
&\quad + (i) \frac{(\Gv -\Gv')}{\Bar{n}} \int_{x}^{a}
    n(x')^3 e^{i(\Gv-\Gv')\cdot\xiv(x')} d{x'} \Biggr] .
\label{pauli_pot1d}
\end{align}

\section{Derivative discontinuity and the HOMO-LUMO gap}\label{derivative_disc}
In the previous section, we have obtained formulas to calculate the functional derivative of $T_s[n]$ at constant particle number.  This means that the variation of $n$ ($\delta n$) is constrained to obey $\int \delta n(\rv)d\rv =0$.  Naturally, this choice produces a potential that is defined up to an arbitrary constant but otherwise varies smoothly with density.

Now we want to address the calculation of discontinuities in the functional derivative as a function of the particle number, and in so doing we will show that the discontinuity itself is a functional of the density.

Let us recall that the particle number is given by
\be\label{NumberConstraint}
N=N_\Omega \int_\Omega n_\Omega(\rv)d\rv \equiv \nu\,.
\ee
The fraction $\nu \equiv N/N_\Omega$ is the number of electrons per unit cell, also known as the filling factor.  In the limit of large $N_\Omega$  the {\it filling factor}  varies over an essentially continuous range of values.  In practice, a relatively small value of $N_\Omega$ (in the range $10-100$)  should be sufficient to simulate an infinitely periodic system.

 The figure below shows schematically the structure of the density space. Each sheet corresponds to a different value of the total particle number.  $N_\Omega$ is large and fixed.  Thus, the change in the filling factor from one sheet to the next is small and becomes infinitesimal in the limit $M \to \infty$. There are no densities ``in-between" the sheets of constant $N$. This reflects the physical fact that the number of electrons is always an integer, even though the filling factor can be fractional.

\begin{figure}[ht]
\begin{center}
\includegraphics[width=8cm]{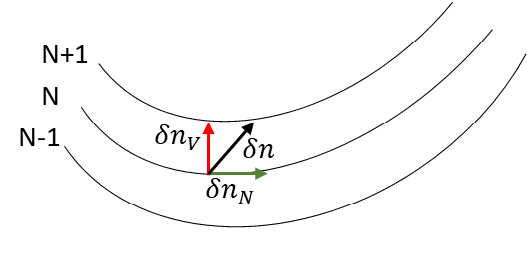}
\caption{Structure of density space  and decomposition of density increment $\delta n$  from $N$- to $N+1$-particle sheets into constant $N$ and constant $V$ components, $\delta n_N$ and $\delta n_V$ respectively. Number of unit cells, $N_\Omega$, is assumed to be large and constant.}
\end{center}
\label{fig:DensitySpace}
\end{figure}

Starting from a  density $n(\rv)$, which satisfies  the condition~(\ref{NumberConstraint}), our algorithm produces the kinetic energy $T_s[n]$ and its functional derivative on the constant-$N$ sheet, which we denote by $V_s[n](\rv)$:
\be
n(\rv) \to \left\{T_s[n],V[n](\rv)\right\}\,.
\ee
We note that the potential $V[n](\rv)$ uniquely determines, via the solution of the Schr\"odinger equation,   a complete set of orthonormal orbitals $\psi_{i\kv}[n](\rv)$ with eigenvalues $\epsilon_{i\kv}[n]$, yielding the density $n(\rv)$ when the $N$  lowest energy orbitals are occupied. On the other hand the potential itself and the eigenvalues $\epsilon_{i\kv}[n]$ are determined only up to a constant, which must be fixed by some supplementary condition, e.g., requiring that $V_s$ has zero average in the unit cell.  However, we {\it cannot} assume that the additional constant that is required to satisfy the supplementary condition varies smoothly with particle number.

To compare functional derivatives at different values of $N$, we must introduce a {\it connection}, which tells us how to ``transport" a density from the N-particle sheet to the $N\pm1$-particle sheet.
The natural way to make this connection is to keep the potential constant.   For example, to connect the $N$-particle sheet to the $N+1$-particle sheet we add a particle to the lowest unoccupied molecular orbital (LUMO), denoted by $\psi_L(\rv)$,  in the potential $V_s[n](\rv)$, which is uniquely associated with $n$ in particle number $N$.  The ``transported density" at particle number $N+1$ is then given by
\be
n_{+}(\rv)=n(\rv)+\frac{1}{N_\Omega}|\psi_L[n](\rv)|^2
\ee
 Similarly,  to connect the $N$-particle sheet to the $N-1$-particle sheet we remove a
 particle from the highest occupied molecular orbital (HOMO), denoted by $\psi_H(\rv)$,  in the potential $V_s[n](\rv)$ and we get
\be
 n_{-}(\rv)=n(\rv)-\frac{1}{N_\Omega}|\psi_H[n](\rv)|^2\,.
\ee
Now according to the variational principle of DFT, we have the following equations
\ber \label{VariationalEquations}
&&\left.\frac{\delta T_s[n]}{\delta n(\rv)}\right\vert_{n,+}= -V[n](\rv)+\mu_+\nn\\
&&\left.\frac{\delta T_s[n]}{\delta n(\rv)}\right\vert_{n,-}=-V[n](\rv)+\mu_-\,,
\eer
where the subscript $+$ means that the derivative is taken along the ``direction" defined by the square of the LUMO orbital $|\psi_L[n](\rv)|^2$, thus increasing the particle number, while the subscript $-$ means that the derivative is taken along the ``direction" defined by the negative of the square of the HOMO orbital $|\psi_L[n](\rv)|^2$, thus decreasing the particle number.  $\mu_+$ and $\mu_-$ are as yet undetermined constants and
the difference $\mu_+-\mu_-$ is precisely the sought discontinuity
\be
\left.\frac{\delta T_s[n]}{\delta n(\rv)}\right\vert_{n,+}-
\left.\frac{\delta T_s[n]}{\delta n(\rv)}\right\vert_{n,-} =\mu_+-\mu_-
\ee

To calculate $\mu_+$ and $\mu_-$ we multiply the first of equations~\ref{VariationalEquations} by $\frac{1}{N_\Omega}|\psi_L[n](\rv)|^2$ and the second by $\frac{1}{N_\Omega}|\psi_H[n](\rv)|^2$ and integrate over $\rv$.  From the definition of the functional derivatives, it follows that
\be
\frac{1}{N_\Omega}\int\left.\frac{\delta T_s[n]}{\delta n(\rv)}\right\vert_{n,+} |\psi_L[n](\rv)|^2 d\rv=T_L[n]
\ee
and
\be
\frac{1}{N_\Omega}\int\left.\frac{\delta T_s[n]}{\delta n(\rv)}\right\vert_{n,-} |\psi_H[n](\rv)|^2 d\rv=T_H[n]
\ee
where $T_L[n]$ and $T_H[n]$ are, respectively,  the kinetic energies associated with the LUMO and HOMO orbitals at density $n$. Similarly, we have
\be
\frac{1}{N_\Omega}\int V(\rv)|\psi_{L(H)}[n](\rv)|^2 d\rv=V_{L(H)}[n]\,,
\ee
where $V_L[n]$ and $V_H[n]$ are the expectation values of $V(\rv)$ in $\psi_L[n]$ and $\psi_H[n]$ respectively. Thus we arrive at
\ber
T_L[n] &=& -V_L[n]+\mu_+\nn\\
T_H[n] &=& -V_H[n]+\mu_-\,.
\eer
Noting that $T_L[n]+V_L[n]=\epsilon_L[n]$ and $T_R[n]+V_R[n]=\epsilon_R[n]$, where $\epsilon_L[n]$ and $\epsilon_R[n]$ are respectively the eigenvalues of the LUMO and the HOMO associated with the potential $V[n]$, we conclude that
\be
\mu_+-\mu_- = \epsilon_L[n]-\epsilon_H[n]\,,
\ee
i.e., the discontinuity of the functional derivative is precisely the HOMO-LUMO gap at the given density.

In the limit of large $N_\Omega$ the discontinuity tends to zero if the filling factor is fractional because $t_L$ and $t_H$ are associated with two infinitesimally close Bloch functions in the same band. But the discontinuity remains finite when the filling factor is integer because in this case the states $\psi_L$ and $\psi_H$ belong to different bands and their eigenvalues differ by the so-called HOMO-LUMO gap.

The calculation of $\epsilon_L[n]$ and $\epsilon_H[n]$ is straightforward once the optimal coefficients $\Bar{\bf C}$ have been determined.  For $\psi_L$ we simply seek the orbital that minimizes the expectation value of $\hat T+\hat V$ and is orthogonal to all the occupied orbitals.  For $\psi_H$ we seek the orbital that maximizes the expectation value of $\hat T+\hat V$ and is a linear combination of the occupied orbitals.  In one dimension the task is greatly facilitated by the fact that we know a priori the occupation numbers $N_k$ and the $k$ values of the HOMO and the LUMO as explained in the next section.

\section{Occupation numbers}\label{occupation_mask}

A potentially crucial issue is how to determine the occupations of the states in momentum space.  In general, the distribution of the occupations $N_{\kv}$ over the  Brillouin zone should be determined by energy minimization.  Suppose for example that we have one electron per unit cell.  The simplest possibility is to search for a solution in which there is exactly one occupied orbital for each $\tilde \kv$ in the BZ.  However in principle, one could also look for a solution in which there are two occupied orbitals for some $\tilde \kv$ and therefore necessarily no occupied orbitals at some other $\tilde\kv$ in the BZ.  In general, we must introduce an integer-valued distribution function $N_{\tilde\kv}$ with average value $\tilde N$ (the number of electrons per unit cell).  Then the energy must be minimized not only with respect to the orbitals but also with respect to their distribution in momentum space.

In one dimension the solution of the ``occupation problem''  is greatly facilitated by known exact features of the band structure\cite{geller1995}.
In this case, there is no overlap between the bands at any given \( k \), and the minima and maxima of each band occur at $k=0$ and $\pi$ respectively for odd bands, or at $k=\pi$ and $0$ respectively for even bands.  Thus, the number of occupied bands for each $k$ can be easily calculated.

We consider for simplicity the case that $N$ and $N_\Omega$ are both odd.  The admissible values of $k$ are
\be
k=\frac{2\pi}{Na}\ell\,,
\ee
where $\ell$ is an integer such that $|\ell|\leq \frac{N_{\Omega}-1}{2}$. We write
\be
N=N_\Omega \bar n+\bar \nu
\ee
where $\bar n$ is the number of fully occupied bands and $\bar \nu$ is the number of electrons in the partially occupied band, of which there is at most one.  Then the occupation number for even $\bar n$ is given by
\ber
N_k &=& \bar n+1\,,~~~~~|\ell|\leq \frac{\bar \nu-1}{2} \nn\\
&=& \bar n\,,~~~~~~~~~~|\ell|> \frac{\bar \nu -1}{2}\,,
\eer
and for odd $\bar n$ by
\ber
N_k &=& \bar n+1\,,~~~~~\frac{N_\Omega-1-\bar \nu}{2} \mathbf{<} |\ell|\leq \frac{N_\Omega-1}{2}\nn\\
&=& \bar n\,,~~~~~~~~~~ {\rm otherwise}\,.
\eer
Similar formulas can be easily obtained for cases when either $N_\Omega$ or $N$ is even, but care must be exerted to resolve the ambiguity of the occupation numbers at the largest or smallest values of $k$ in the partially filled band.

\begin{figure*}[ht]
    \centering
    \begin{minipage}[b]{0.48\textwidth}
        \centering
        \raisebox{10mm}{\includegraphics[width=\textwidth]{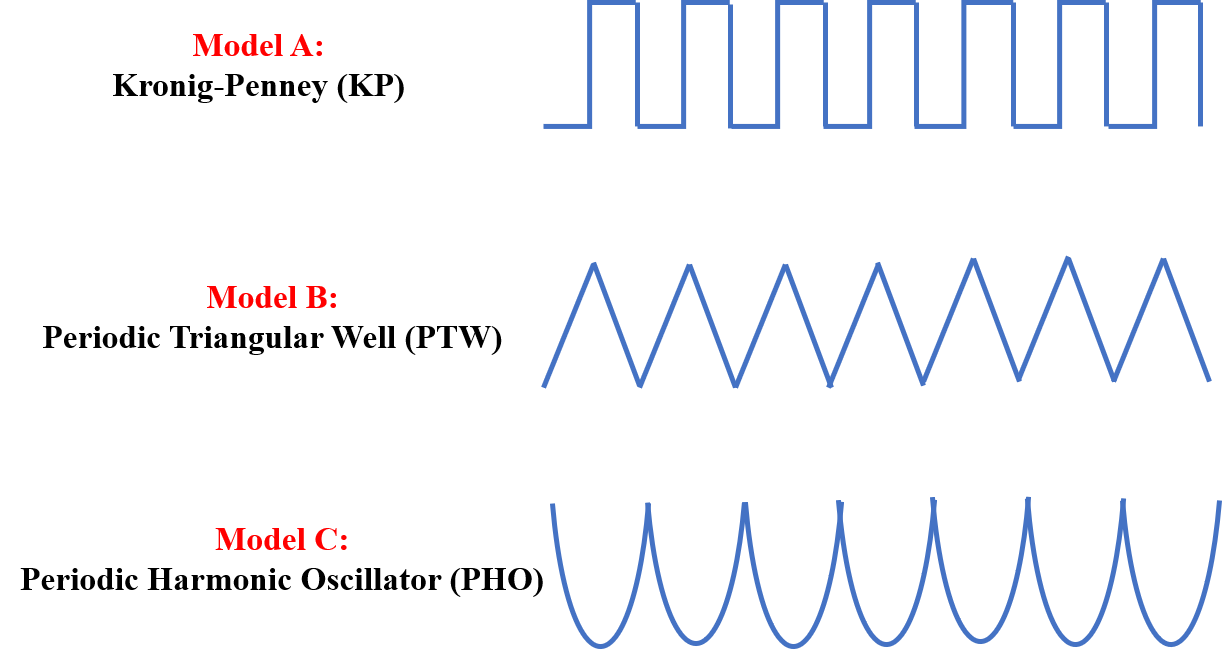}}
        \caption*{(a)}
        \label{fig:subfig1}
    \end{minipage}
    \hfill
    \begin{minipage}[b]{0.48\textwidth}
        \centering
        \includegraphics[width=\textwidth]{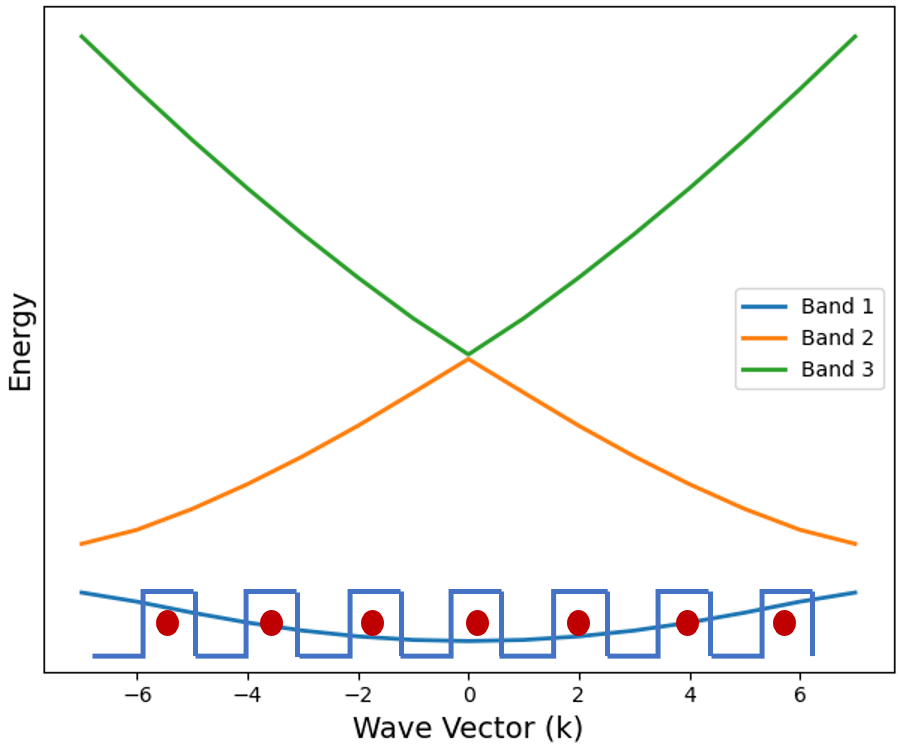}
        \caption*{(b)}
        \label{fig:subfig2}
    \end{minipage}
    \caption{(a) Exactly solvable model potentials used for benchmarking. (b) Schematic representation of the energy band diagram of a one-dimensional periodic potential.}
    \label{fig:modelpot_all}
\end{figure*}

\section{Benchmarking the algorithm}\label{benchmarking}

\begingroup
\begin{table*}[htb]
    \centering
    \caption{Comparison between the value of the Pauli kinetic energy ($T_{P}\big|_{\text{cs}}$) obtained from our constrained search algorithm with the exact value ($T_{P}\big|_{\text{exact}}$) computed from the solution of the Schr\"odinger equation for different model potentials. All the energies are in units of $\frac{\hbar^2}{ma^2}={\rm Hartree} \times \left(\frac{a_B}{a}\right)^2$, where $a_B$ is the Bohr radius. Absolute value of the  relative error is denoted by, $|\Delta| = \frac{\left \vert T_{P}\big|_{\text{cs}} - T_{P}\big|_{\text{exact}}\right \vert}{T_{P}\big|_{\text{exact}}}$. Also shown, for reference, the total kinetic energy in the Thomas-Fermi approximation ($T_{F}\big|_{\text{exact}}$) and the exact Bosonic energy ($T_{B}\big|_{\text{exact}}$) from Eq.~(\ref{TBosonic}).}
    \label{tab:accuracy_data}
    \begin{adjustbox} {width=0.80\textwidth}
    {\fontsize{3}{4}\selectfont 
    \begin{tabular}{|c|c|c|c|c|c|c|}
        \hline
        \multirow{2}{*}{} & \multirow{2}{*}{} & \multicolumn{5}{c|}{} \\
        \multirow{2}{*}{$(N, N_{\Omega})$} & \multirow{2}{*}{$\nu$} & \multicolumn{5}{c|}{\textbf{Model A (KP)}} \\
        \cline{3-7}
          & & $T_{F}\big|_{\text{exact}}$ & $T_{B}\big|_{\text{exact}}$ & $T_{P}\big|_{\text{exact}}$ & $T_{P}\big|_{\text{cs}}$ & $|\Delta|$ \\
         & & & & & & \\
        \hline
        & & & & & & \\
        (15,15) & 1    & 2.3047  & 0.6759 & 1.3144  & 1.3142 & 1.5$\times$ 10$^{\text{-4}}$ \\
        (22,15) & 1.46 & 3.7109  & 0.0843 & 3.6141  & 3.6139 & 6$\times$ 10$^{\text{-5}}$ \\
        (30,15) & 2    & 6.6881  & 0.0466 & 6.6292  & 6.6282 & 1.5$\times$ 10$^{\text{-4}}$ \\
        (37,15) & 2.46 & 10.0607 & 0.0149 & 10.0363 & 10.0358 & 5$\times$ 10$^{\text{-5}}$ \\
        (45,15) & 3    & 14.8606 & 0.0295 & 14.8066 & 14.8059 & 5$\times$ 10$^{\text{-5}}$ \\
        & & & & & & \\
        \hline
        \multirow{2}{*}{} & \multirow{2}{*}{} & \multicolumn{5}{c|}{} \\
        \multirow{2}{*}{$(N, N_{\Omega})$} & \multirow{2}{*}{$\nu$} & \multicolumn{5}{c|}{\textbf{Model B (PTW)}} \\
        \cline{3-7}
         & & $T_{F}\big|_{\text{exact}}$ & $T_{B}\big|_{\text{exact}}$ & $T_{P}\big|_{\text{exact}}$ & $T_{P}\big|_{\text{cs}}$ & $|\Delta|$ \\
        & & & & & & \\
        \hline
        & & & & & & \\
        (15,15) & 1   & 1.9935 & 0.3515 & 1.4560 & 1.4555 & 3.4$\times$ 10$^{\text{-4}}$ \\
        (22,15) & 1.46 & 3.6099  & 0.0333 & 3.5782  & 3.5781  & 3$\times$ 10$^{\text{-5}}$ \\
        (30,15) & 2    & 6.6266  & 0.0216 & 6.6030  & 6.6027  & 5$\times$ 10$^{\text{-5}}$ \\
        (37,15) & 2.46 & 10.0279 & 0.0034 & 10.0160 & 10.0157 & 3$\times$ 10$^{\text{-5}}$ \\
        (45,15) & 3    & 14.8180 & 0.0025 & 14.8069 & 14.8063 & 4$\times$ 10$^{\text{-5}}$ \\
         & & & & & & \\
        \hline
        \multirow{2}{*}{} & \multirow{2}{*}{} & \multicolumn{5}{c|}{} \\
        \multirow{2}{*}{$(N, N_{\Omega})$} & \multirow{2}{*}{$\nu$} & \multicolumn{5}{c|}{\textbf{Model C (PHO)}} \\
        \cline{3-7}
         & & $T_{F}\big|_{\text{exact}}$ & $T_{B}\big|_{\text{exact}}$ & $T_{P}\big|_{\text{exact}}$ & $T_{P}\big|_{\text{cs}}$ & $|\Delta|$ \\
         & & & & & & \\
        \hline 
        & & & & & & \\
        (15,15) & 1    & 1.9775  & 0.3428 & 1.4592 & 1.4586 & 4.1$\times$ 10$^{\text{-4}}$ \\
        (22,15) & 1.46 & 3.6086  & 0.0351 & 3.5771  & 3.5770  & 2.8$\times$ 10$^{\text{-5}}$ \\
        (30,15) & 2    & 6.6511  & 0.0463 & 6.5825  & 6.5822  & 4.5$\times$ 10$^{\text{-5}}$ \\
        (37,15) & 2.46 & 10.0303 & 0.0050 & 10.0162 & 10.0158 & 3.9$\times$ 10$^{\text{-5}}$ \\
        (45,15) & 3    & 14.8195 & 0.0033 & 14.8072 & 14.8067 & 3.3$\times$ 10$^{\text{-5}}$ \\
        & & & & & & \\
        \hline
    \end{tabular}
    }
    \end{adjustbox}
\end{table*}
\endgroup

To benchmark our model we will consider three exactly solvable models with $N$ electrons distributed over $N_\Omega$ unit cells: (i) the Kronig-Penney (KP) model (ii) the periodic triangular well (PTW), and (iii) the periodic harmonic oscillator well (PHO).
These three model potentials and the qualitative nature of their band structures are shown in Fig.~\ref{fig:modelpot_all}.
The KP model is analytically solvable, while the PTW and PHO models are solved numerically to generate the dataset for comparison.

The constrained search algorithm is implemented with several key parameters to ensure accuracy and efficiency in the optimization process. The Adam optimizer \cite{kinga2015} is used with a learning rate of \(10^{-3}\). Convergence is achieved when the difference in the loss function between 50 consecutive iterations falls below a threshold value, set to \(10^{-4}\) for our calculations. The damping coefficients \(c_Q\) (as defined in Eq.~\ref{damping_coeff}) are also specified. The basis set size (number of G vectors) is set to 100. Numerical integrations are performed on an equally spaced grid, with a particularly high number of grid points, approximately 25000, used to achieve the required accuracy in energy calculations. Additionally, the orbital coefficients ${\bf C}$ are initialized by occupying the  G-vectors of smaller magnitude first, which aids in accelerating the convergence of the optimization procedure. These parameters together optimize the algorithm's efficiency and ensure the precision of the results.

\begin{figure*}[htb]  
    \centering
    \includegraphics[width=0.99\textwidth]{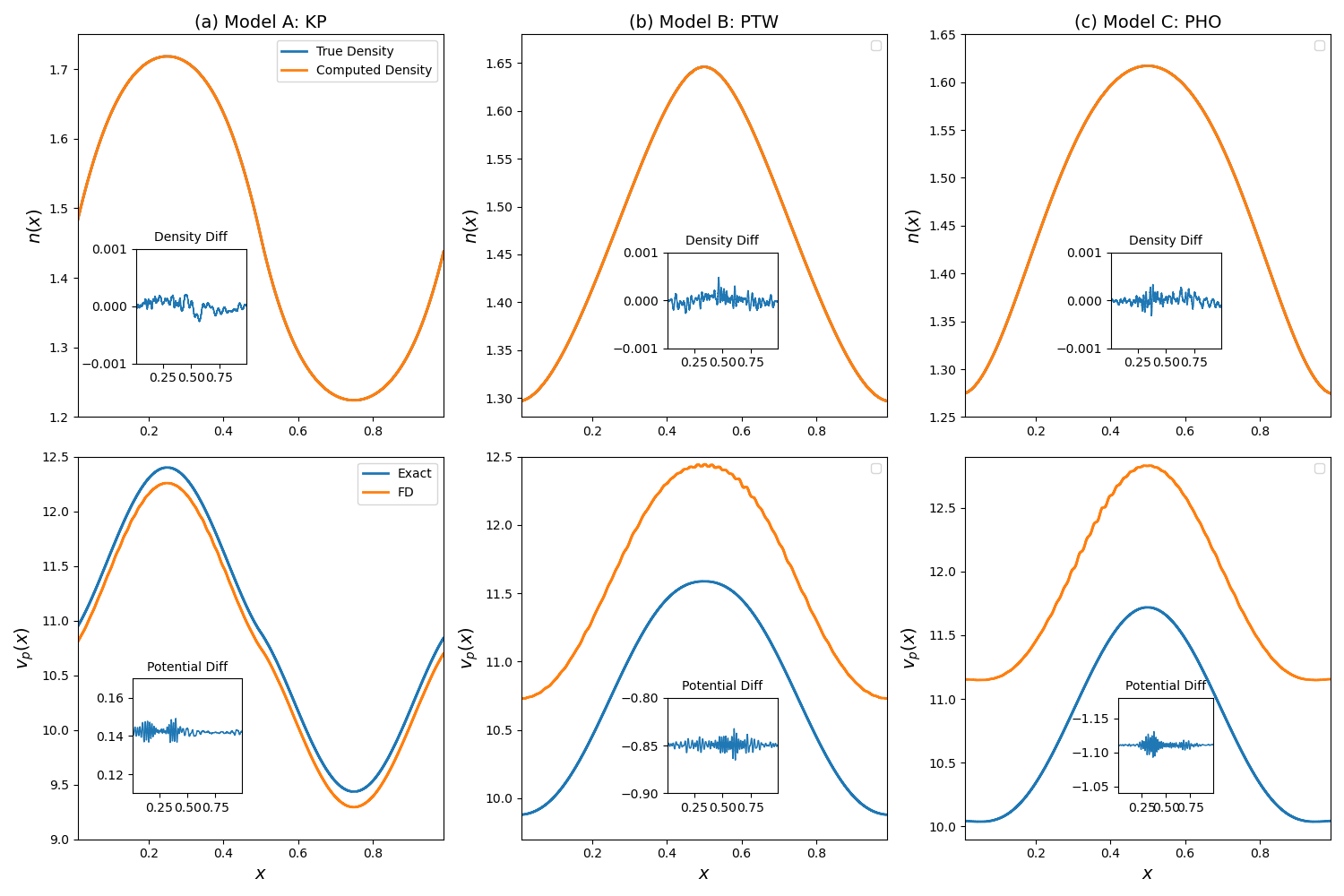}  
    \caption{Benchmarking the algorithm: comparison of the true and computed densities from the constrained search method (top row); the inset plots show the difference between the exact and the computed densities. The bottom row presents an analysis of the exact Pauli potential (from Eq.~\ref{exact_pauli}) against those obtained via the constrained search approach with the functional derivative method (from Eq.~\ref{pauli_pot1d}) for $\nu = 1.46 (N=22, N_{\Omega}=15)$ across three of our model potentials. The inset plots show the difference between the exact and the computed Pauli potentials for each case.}
    \label{fig:ts_vp_accuracy}
\end{figure*}

\subsection{Kinetic energy}
We begin by comparing the numerical results of $T_P[n]$ obtained from our constrained search approach with exact values calculated for the three different model potentials described in Fig.~\ref{fig:modelpot_all}. The comparison is presented in Table~\ref{tab:accuracy_data}.  In each section of the Table we fix the parameters \((N, N_{\Omega})\) and the filling factor ($\nu = \frac{N}{N_{\Omega}}$), and report the following energy values: Thomas-Fermi kinetic energy (\(T_F\)), Bosonic energy (\(T_B\)), and Pauli energy (\(T_P\)), along with the calculated Pauli energy (\(T_P\vert_{\text{cs}}\)) and the absolute value of the relative error \(|\Delta|\), given by  \[
\left| \frac{T_{P}\vert_{\text{cs}} - T_{P}\vert_{\text{exact}}}{T_{P}\vert_{\text{exact}}} \right|.
\]
A rigorous  inequality connecting these energies  is easily derived \footnote{Here $F[N] \equiv 1 + \frac{2-3\text{mod}[N, 2]}{N^2}$. For large N, $F[N] \simeq 1. \text{ This is obtained by taking the sum of the indices of the state.} \break \text{These indices are determined by populating the states to find the optimal slater determinant with minimum energy.}$}:
\begin{equation}
    T_s[n] \leq T_B[n]
    + F[N] T_F[n]\,,
\end{equation}
which can be  simplified to
\begin{equation}
    T_P[n] \leq F[N] T_F[n].
\end{equation}
For all the cases tabulated here, these bounds are well satisfied. The results show consistently low relative errors ($|\Delta|$), all below $1 \times 10^{-4}$.
The method performs well across different potentials without a significant trend in error variation with increasing \((N, \nu)\). This confirms its robustness and reliability for accurately computing the Pauli kinetic energy functional.

\subsection{Pauli potential}
In Fig.~\ref{fig:ts_vp_accuracy} we visually compare the computed and exact densities, as well as the computed and exact Pauli potentials for our three models. The comparison is done for  $N = 22, N_{\Omega} = 15$ ($\nu = 1.46$), consistent with the data presented in  Table~\ref{tab:accuracy_data}.
The \textit{exact} Pauli potentials are calculated from the well known formula \cite{levy1988}
\be\label{exact_pauli}
V_{P}^{\text{exact}}(x) = \frac{1}{n(x)} \sum_{i=1}^{N} (\varepsilon_N - \varepsilon_i) n_i(x)
- \frac{1}{2n(x)} \sum_{i=1}^{N} \varphi_i^*(x) \nabla^2 \varphi_i(x)
+ \frac{\nabla^2 \sqrt{n(x)}}{2\sqrt{n(x)}}\,,
\ee
where the sums run over the occupied states, $\varphi_i(x)$  obtained from the exact diagonalization of the Hamiltonian of our three model potentials.  Notice that this formula contains the HOMO eigenvalue $\epsilon_{N}$. In a finite system, this would ensure that the Pauli potential tends to zero at infinity.

The top row in each panel compares the exact and the computed densities, while the bottom row shows the exact Pauli potential alongside the Pauli potential computed from the functional derivative (FD) with respect to density at constant particle number. Insets in each plot depict the differences between exact and computed quantities. The density differences indicate that deviations remain within $\pm 0.001$, demonstrating high numerical precision. Similarly, the potential differences are in the range of $\pm 0.02$ to $\pm 0.05$ in all three cases.

In the lower part of Fig.~\ref{fig:ts_vp_accuracy}  a key observation is that while the Pauli potential \textit{computed}  from Eq.~\ref{pauli_pot1d} closely follows the shape of the exact potential, it exhibits a uniform shift. This shift does not affect the functional derivative's ability to capture the essential features and variations of the Pauli potential but introduces an offset in absolute values. Since only potential differences influence physical observables, such shifts do not undermine the validity of the method.

\subsection{Bosonic potential and full potential}
\begin{figure*}[h]  
    \centering
    \begin{subfigure}[b]{0.99\textwidth}
        \centering
        \includegraphics[width=\textwidth]{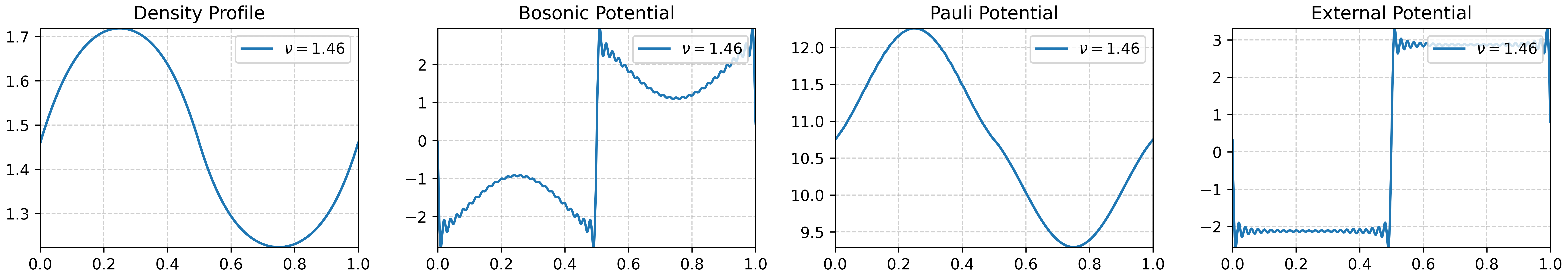}  
        \label{fig:subfig1}
    \end{subfigure}

    \begin{subfigure}[b]{0.99\textwidth}
        \centering
        \includegraphics[width=\textwidth]{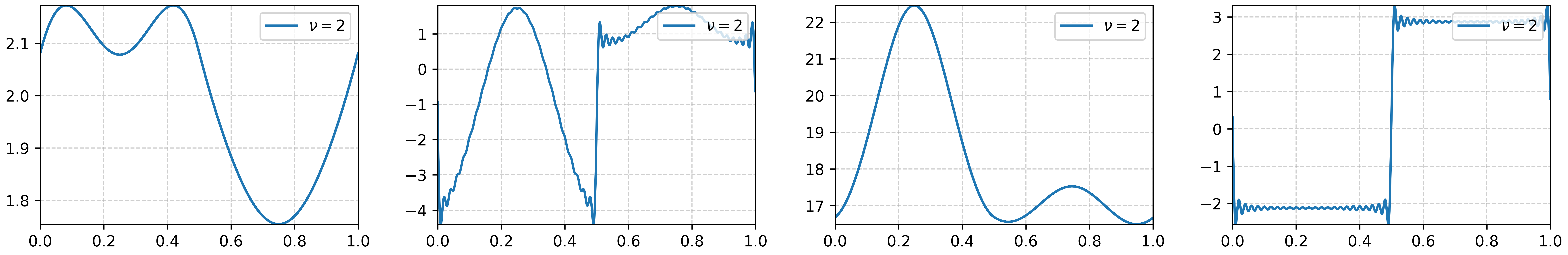}  
        \label{fig:subfig2}
    \end{subfigure}

    \begin{subfigure}[b]{0.99\textwidth}
        \centering
        \includegraphics[width=\textwidth]{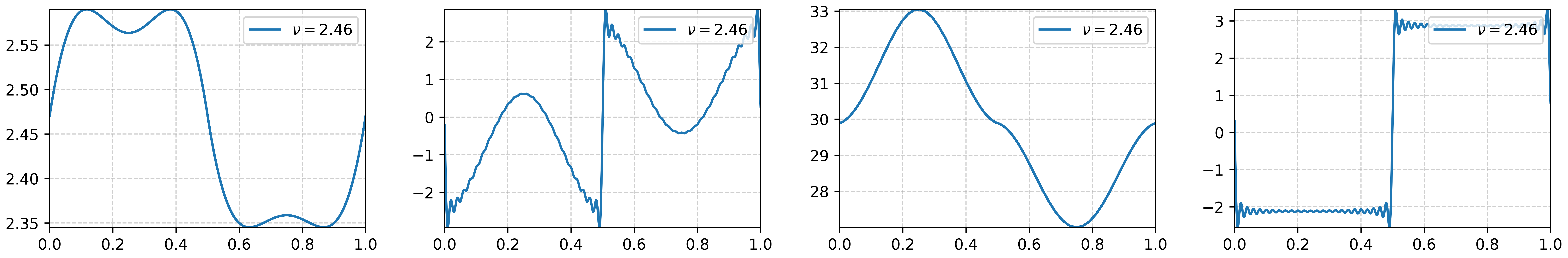}  
        \label{fig:subfig3}
    \end{subfigure}

    \caption{Steps of the density to potential inversion for three ground state densities of the KP model potential with $N_\Omega=15$, shown on the leftmost column. The Bosonic potential is computed from Eq.~(\ref{TBosonic}), the Pauli potential from our constrained search algorithm. The external potential (a step function for the KP model) is calculated from $V_B$ and $V_P$ according to Eq.~(\ref{potcomp}). The average of the potential is set to zero within the unit cell.}
    \label{fig:density_pot_inver}
\end{figure*}

Figure \ref{fig:density_pot_inver} presents a detailed schematic representation of the density-to-potential inversion process, illustrating the decomposition of the total potential into its bosonic and Pauli components across different density profiles. The results are shown for three representative values of the filling factor, $\nu = 1.46, 2 , 2.46$,
arranged in rows. Each row consists of four subfigures corresponding to the density profile, bosonic potential, Pauli potential, and external potential.

The leftmost column shows the density profiles calculated for different values of $\nu$ that capture the expected spatial variations and are smooth functions. Importantly, these density profiles serve as the starting point for the inversion process, where the goal is to reconstruct the underlying effective potentials that generate them.

The second and third columns display the bosonic potential and Pauli potential, respectively. These potentials exhibit distinct characteristics, highlighting their respective contributions to the total effective potential. The bosonic potential, which primarily captures interaction effects, shows variations with sharper features and discontinuities at certain points, especially for lower $\nu$. The Pauli potential, arising from exchange and correlation effects in the system, exhibits a smoother profile with significant peak structures. Notably, as the density increases (from $\nu = 1.46$ to $\nu = 2.46$), the Pauli potential undergoes a substantial increase in magnitude, reflecting stronger fermionic effects.

The final column presents the external potential required to maintain the given density profile.
The external potential \(V_{S}(x)\) (generated from analytical functional derivatives) is defined as the difference between the Bosonic potential and the Pauli potential following Eq.~\ref{potcomp}.
As expected, the external potential remains constant and equal to the KP potential in this case. The successful inversion process is evident from the consistency of the reconstructed potentials across different density regimes.

\subsection{Derivative discontinuity}
In this section, we examine the  non-analytic dependence of the kinetic energy functional on the particle number, which results in the well-known \textit{derivative discontinuity} at integer particle numbers.
\begin{figure*}[htb]  
    \centering
    \includegraphics[width=0.9\textwidth]{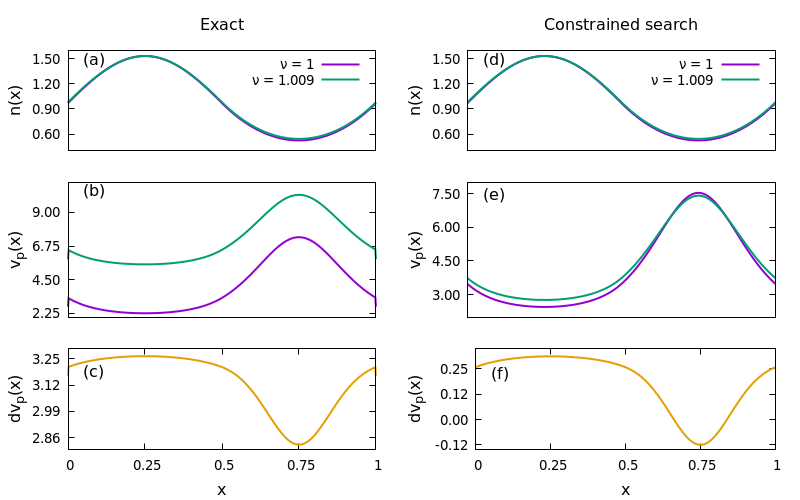}  
    \caption{(a) Exact ground state densities of the KP model with $N_\Omega=101$ for $\nu=1$ ($N=101$) and $\nu=1.009$ ($N=102$). (b) Exact Pauli potentials, calculated according to Eq.~\ref{exact_pauli} for the same values of $N$ and $\nu$ as in (a). (c) Difference between exact Pauli potentials at $\nu=1.009$ and $\nu=1$. This difference is not exactly constant in space but tends to a constant in the limit $N_\Omega \to \infty$.\cite{kraisler2020} Panels (d)--(f) show the results of the constrained search algorithm for sthe ame quantities plotted in panels (a)--(c) respectively. Notice that the Pauli potential in panel (e) has no discontinuity, being given as an explicit functional of density (Eq.~(\ref{pauli_pot1d})) which remains almost unchanged in going from $\nu=1$ to $\nu=1.009$.
    Densities are in units of $a^{-1}$ and energies are in units of $\hbar^2/ma^2$.}
    \label{fig:derv_disc}
\end{figure*}
To do this, in Fig.~\ref{fig:derv_disc} we compare the densities and the Pauli potentials calculated for two closely spaced filling factors, \(\nu = 1\) and \(\nu = 1.009\), corresponding to particle numbers \(N = 101\) and \(N = 102\), in the KP model. On the left side of  Fig.~\ref{fig:derv_disc} (panels (a)-(c)) the calculations are done exactly, i.e., diagonalizing the KP Hamiltonian and using Eq.~\ref{exact_pauli} for the Pauli potential.  On the right side (panels (d)-(f))  we use instead the constrained search algorithm, starting from the density and computing the Pauli potential as a functional derivative with respect to the density at constant particle number, i.e., via Eq.~\ref{pauli_pot1d}.    In panels (c) and (f)  we plot the difference $dv_p(x)$ between the two Pauli potentials calculated at the two nearby filling factors. The striking result in panel (b)  is that although the density remains almost unchanged in going from \(\nu = 1\) to \(\nu = 1.009\), the Pauli potential from Eq.~\ref{exact_pauli} exhibits a discontinuity as the particle number changes from $N=101$ to $N=102$. This is the \textit{derivative discontinuity}\cite{kraisler2020}. Notice that, $dv_p(x)$ (panel (c))  is not exactly constant in space, but it becomes constant in the limit  \(N_\Omega \to \infty\) \cite{levy2014, kraisler2020}. This crucial feature is missing in the Pauli potential calculated from the functional derivative of the kinetic energy functional in the constrained search approach - see panel (e). The discrepancy arises because the functional derivative (Eq.~\ref{pauli_pot1d}) in (e) is calculated at a fixed particle number and hence misses crucial information about the change in particle number, which, when properly handled,  leads to the discontinuity of the Pauli potential in panel (b).

In fact, as discussed in Section~\ref{derivative_disc}, the jump in the Pauli potential at integer particle numbers is fully determined by the HOMO-LUMO gap, which can be calculated by the constrained search algorithm. Table \ref{tab:homo_lumo_data} benchmarks the accuracy of the constrained search method in determining the HOMO-LUMO energy gap ($\epsilon_g$) by comparing the numerical results with exact values for different model potentials. The table presents the highest occupied molecular orbital (HOMO) energy ($\epsilon_H$), the lowest unoccupied molecular orbital (LUMO) energy ($\epsilon_L$), the computed HOMO-LUMO gap ($\epsilon_g$), and the absolute value of the relative error in $\epsilon_g$. The values in parentheses, except in the first column, indicate reference results obtained from diagonalization (referred to as exact) of the KP Hamiltonian.
\begingroup
\begin{table*}
\centering
\caption{Comparison between the HOMO-LUMO gaps $\epsilon_g=\epsilon_L-\epsilon_H$ computed from the constrained search algorithm and the exact ones obtained from diagonalization for our three model potentials.  The exact values are in parenthesis. All the energies are in units of $\frac{\hbar^2}{ma^2}={\rm Hartree} \times \left(\frac{a_B}{a}\right)^2$, where $a_B$ is the Bohr radius. Absolute value of the relative error is denoted by, $|\Delta| = \frac{\left \vert \epsilon_{g}\big|_{\text{cs}} - \epsilon_{g}\big|_{\text{diagonalization}}\right \vert}{\epsilon_{g}\big|_{\text{diagonalization}}}$.}
\begin{adjustbox} {width=0.99\textwidth}
{\fontsize{6}{7}\selectfont 
    \begin{tabular}{|c|c|c|c|c|c|c|c|c|c|c|c|c|}
    \hline
    & \multicolumn{4}{c|}{} & \multicolumn{4}{c|}{} & \multicolumn{4}{c|}{} \\
     & \multicolumn{4}{c|}{\text{Model A}} & \multicolumn{4}{c|}{\text{Model B}} & \multicolumn{4}{c|}{\text{Model C}} \\
     \textbf{($N_{\Omega}$, N)} & \multicolumn{4}{c|}{KP} & \multicolumn{4}{c|}{PTW} & \multicolumn{4}{c|}{PHO} \\
    \cline{2-13}
     & & & & & & & & & & & &\\
     & $\epsilon_{\text{H}}$ & $\epsilon_{\text{L}}$ & $\epsilon_{g}$ & $\left| \Delta \right|$ & $\epsilon_{\text{H}}$ & $\epsilon_{\text{L}}$ & $\epsilon_{g}$ & $\left| \Delta \right|$ &  $\epsilon_{\text{H}}$ & $\epsilon_{\text{L}}$ & $\epsilon_{g}$ & $\left| \Delta \right| $\\
     & & & & & & & & & & & &\\
     \hline
           &   &   &   &   &   & & & & & & &\\
     (51,51) & 5.5647  & 8.7846  & 3.2199  & 0.006  & 1.5936  & 3.6680 & 2.0743 & 0.005 & 0.5895 & 2.6345 & 2.0449 & 0.0046 \\
     &   &   & (3.1984)  &   &   & & (2.0630) & & & & (2.0354) &\\
     (101,101) & 5.5794  & 8.7747  & 3.1953  & 0.004  & 1.6181  & 3.6636 & 2.0454 &  0.005 & 0.6144 & 2.6319 & 2.0175 & 0.0048 \\
     &   &   & (3.1808)  &   &   & & (2.0347) & & & & (2.0077) &\\
    \hline
              &   &   &   &   &   &  &  & & & & &\\
     (5, 10)  & 25.4089  &  25.7267 & 0.3178  & 0.014  & 16.7195  & 17.0062 & 0.2867 & 1.995 & 14.7187 & 15.3164 & 0.5977 & 0.003 \\
     (11, 22) & 23.9319  &  24.2582 & 0.3263  & 0.041  & 19.6042  & 19.7098 & 0.1055 & 0.102 & 17.9823 & 18.5908 & 0.6085 & 0.015 \\
     (15, 30) & 23.5715  &  23.8873 & 0.3158  &  0.007 & 20.1644  & 20.2720 & 0.1075 & 0.123 & 17.9823 & 18.5908 & 0.6085 & 0.015 \\
     &   &   & (0.3133)  &   &   & & (0.0957) & & & & (0.5995) &\\
              &   &   &   &   &   &  &  & & & &  &\\
    \hline
                &   &   &   &   &   &  &  & & & & &\\
     (51, 153)  & 45.32166  & 46.9032  & 1.5815  & 0.010  & 44.8237  & 46.0258 & 1.2020 & 0.0164 & 43.6712 & 44.8733 & 1.2021 & 0.0128 \\
                &           &          & (1.5655)  &   &   &  & (1.1825) & & & & (1.1869) &\\
     (101, 303) & 45.6888  & 46.8887  & 1.1999  & 0.002  & 45.3118  & 45.9518 & 0.6400  & 0.0187 & 44.1466 & 44.7978 & 0.6512 & 0.0229 \\
     &   &   & (1.2034)  &   &   & & (0.6282) & & & & (0.6366) &\\
                &   &   &   &   &   &  &  & & & & &\\
    \hline
\end{tabular}
} 
\end{adjustbox}
\label{tab:homo_lumo_data}
\end{table*}
\endgroup

The results suggest that the method captures the HOMO-LUMO gap with reasonable precision.
The consistent trend of computed values slightly deviating from reference results indicates that systematic refinements such as reducing the increment in the fractional filling factor  ($N_\Omega \rightarrow \infty$) could improve accuracy \cite{levy2014, kraisler2020}. When considered alongside the previous benchmarking analyses for kinetic energy and Pauli potentials, This table further validates the method's reliability in calculating the orbitals from the density.

\subsection{Optimized orbitals}

\begin{figure*}[htb]  
    \centering
    \includegraphics[width=0.8\textwidth]{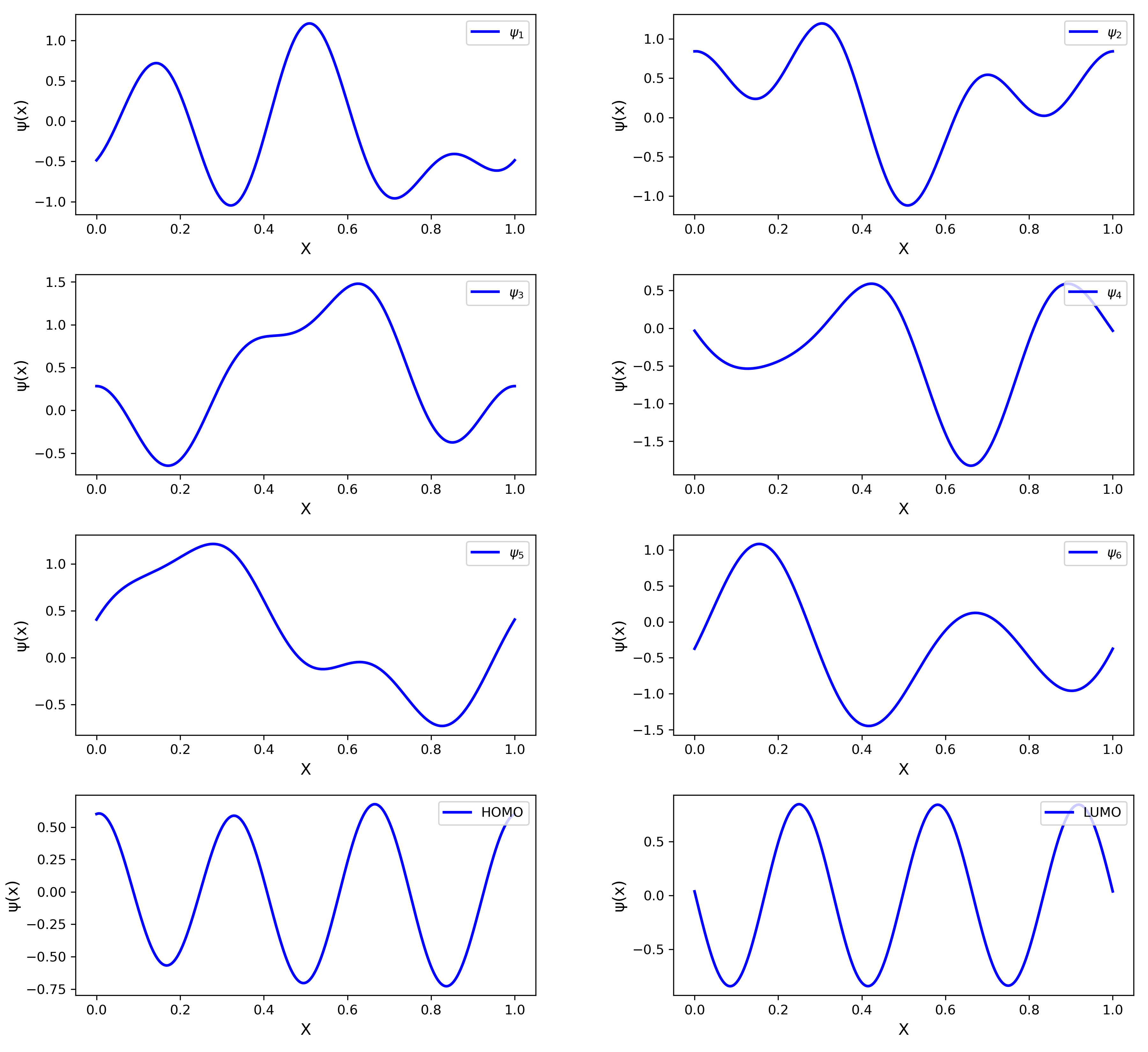}  
    \caption{Top three rows: Plots of the real parts of the 6 occupied orbitals found by the constrained search algorithm applied to the KP ground state density with $N=6$, $N_\Omega=1$.
    Notice that these are not the 6 lowest energy eigenstates of the KP model but they are related to them by a unitary rotation.  Bottom row: Plots of the HOMO and LUMO orbitals for the same model density as above.}
    \label{fig:bloch_wf}
\end{figure*}

Lastly,  we comment on the orbitals that we get from the constrained search algorithm.
As shown in Fig.~\ref{fig:bloch_wf}, upon optimization, we do not obtain canonical Kohn-Sham orbitals but linear combinations of those orbitals, each possessing both real and imaginary components. The nodal structures of the HOMO and LUMO are contingent upon the particle number, N:
\begin{itemize}
    \item For odd \( N \):
    \begin{itemize}
        \item the real part of HOMO exhibits \( N-1 \) nodes.
        \item the real part of LUMO exhibits \( N+1 \) nodes.
    \end{itemize}
    \item For even \( N \):
    \begin{itemize}
        \item the real part of both HOMO and LUMO exhibit \( N \) nodes.
    \end{itemize}
\end{itemize}
The imaginary components of these wavefunctions follow the same nodal patterns as their real counterparts for $N>1$. Consequently, the Bloch states are constrained to possess a maximum number of nodes corresponding to those specified for the eigenstates. As an example, it is shown in the last row of figure~\ref{fig:bloch_wf}, for $N=6$ (an even number of particles) the number of nodes in both real parts of the HOMO and LUMO is equal to N.

\section{Discussion and outlook}\label{outlook}
In summary, we have developed a very accurate algorithm for computing the kinetic energy, the external potential, and the HOMO-LUMO gap that, according to the Hohenberg-Kohn theorem, are uniquely associated with a {\it periodic} density in a one-dimensional system of non-interacting fermions. For typical periodicities on the order, of say, $3 \AA$ our energies, benchmarked against exact calculations in model systems, are well within ``chemical accuracy".

The extension of this algorithm to higher dimensional systems will be, obviously, more time-consuming but it appears to be well within reach. A key ingredient of this extension will be the construction of optimal sets of equidensity orbitals since the solution of the Jacobian problem is highly non-unique in more than one-dimension. Progress in this direction will be reported elsewhere.

An immediate application of our algorithm is the calculation of exact exchange-correlation energies and potentials. If the exact ground state density and energy can be obtained for an interacting many-body system through accurate wave-function methods such as configuration interaction and quantum Monte Carlo, then the exact exchange-correlation energy can be obtained by subtracting from the exact energy of the non-interacting kinetic energy and the Hartree energy. Similarly, the exact exchange-correlation potential can be obtained by subtracting the non-interacting potential and the Hartree potential from the exact external potential.

Looking forward, our algorithm can be used to generate a ``database" of kinetic energies and potentials versus periodic densities, the latter being parameterized as sums of atom-centered positive distributions. It should then be possible to train a neural network on such a data set, effectively producing the universal kinetic energy and potential functional for arbitrary periodic densities of the stated form.

As a final point, we notice that the availability of exact Kohn-Sham orbitals (up to unitary transformation) will allow us to compute the exact exchange energy functional, simply by evaluating a Coulomb integral on the exact one-particle density matrix.

\section{Acknowledgements}
We thank Dr. Kati Finzel for several enlightening discussions on the Pauli potential. We thank Stephen Dale, Tianbo Li, and Min Lin for sharing their expertise on JAX and differentiable programming. This research is supported by the Ministry of Education, Singapore, under its Research Centre of Excellence award to the Institute for Functional Intelligent Materials (I-FIM, project No. EDUNC-33-18-279-V12). ZS is supported by an EBD-IPP scholarship.

\clearpage
\bibliography{ref}
\section{Appendix}
\subsection{Exact implementation of density constraint}
One way to improve the accuracy of the algorithm is to combine the penalty method with a unitary transformation which maps an approximate density $n_a(\rv)$ (obtained by the penalty method) to the target density $n(\rv)$, while preserving the idempotency of the density matrix.\footnote{The approximate density is computed as  $n_a(\rv) = n(\rv) \sum_{\qv}e^{-i\qv\cdot\xiv(\rv)} \frac{1}{N}\sum_{\kv} [C \cdot C^\dagger]_{\kv,\kv+\qv}$ and is expected to be  close to $n(\rv)$ if the penalty method works reasonably well.}
To this end, let us consider the bijective map $\zetav: {\cal B}\to {\cal B}$ defined as the solution of the generalized Jacobian problem\footnote{Here and in the following we take the Jacobian to be the absolute value of the corresponding determinant, i.e., a positive-definite quantity.}
\be\label{JacobianProblem2}
n_a(\zetav(\rv)) \left\vert \frac{\partial \zeta_i(\rv)}{\partial \rv_j}\right\vert=n(\rv)\,,
\ee
with
\be
\zetav(\rv+\Rv)=\zetav(\rv)+\Rv\,.
\ee
This map induces a bijective map on the Hilbert space of one-particle wave function ($D_{\zetav}:{\cal H}_1 \to {\cal H}_1$) such that
\be\label{DensityMap}
[D_{\zetav} \psi](\rv) = \sqrt{\frac{n(\rv)}{n_a(\zetav(\rv))}} \psi(\zetav(\rv))
\ee
A completely analogous transformation is induced in the Hilbert space of $N$-particle wave functions ($D_N:{\cal H}_N \to {\cal H}_N$), where $D_N = \otimes_{i=1}^N D(i)$ where the tensor product runs over the particles.  The proof is simple.   First of all it is clear that if $|\psi(\rv)|^2 = n_a(\rv)$ then $|[D \psi](\rv)|^2=n(\rv)$ as required.  Second, the scalar product of two wave functions is preserved by the transformation:
\ber
\int_{\cal B} d\rv  [D\psi_1]^*(\rv) [D\psi_2](\rv)  &=& \int_{\cal B} d\rv \frac{n(\rv)}{n_a(\zetav(\rv))} \psi_1^*(\zetav(\rv))\psi_2(\zetav(\rv))\nn\\
&=& \int_{\cal B}  d\zetav ~\psi_1^*(\zetav)\psi_2(\zetav)\,.
\eer
Thus, the transformation is unitary.  The  transformation can be performed directly on the coefficients $C$, using the unitary matrix
\be
U_{\kv,\kv'}=\frac{1}{N}\int_{\cal B}d\rv~n(\rv)  \sqrt{\frac{n(\zetav(\rv))}{n_a(\zetav(\rv))}}e^{-i\kv\cdot\xiv(\rv)} e^{i\kv'\cdot\xiv(\zetav(\rv))}\,,
\ee
so that the initial coefficients $C_{i,\kv}$, which gave the approximate density $n_a(\rv)$, are replaced by $\sum_{\kv'}U_{\kv,\kv'}C_{i,\kv'}$, which give exactly the target density $n(\rv)$.

\end{document}